\documentclass[longauth]{aa}  

\usepackage{graphicx}
\usepackage{txfonts}
\usepackage{lipsum}
\usepackage{subcaption}         
\usepackage{lscape}             
\usepackage{placeins}           
\usepackage{hyperref}
\hypersetup{colorlinks=true, citecolor=blue, linkcolor=blue}

\usepackage{multirow}

\newcommand{\civ}{C{\,\tiny IV}}
\newcommand{\mgii}{Mg{\,\tiny II}}
\newcommand{\oi}{O{\,\tiny I}}
\newcommand{\oiii}{[O{\,\tiny III}]}
\newcommand{\feii}{Fe{\,\tiny II}}
\newcommand{\hi}{H{\,\tiny I}}
\newcommand{\hb}{H$\beta$}
\newcommand{\hg}{H$\gamma$}
\newcommand{\rfe}{\mathcal{R}_\mathrm{Fe}}

\begin{document}

   \title{Spatially resolved broad line region in a quasar at z=4}
   \subtitle{Dynamical black hole mass and prominent outflow}

    \author{
    GRAVITY+ Collaboration\thanks{Corresponding authors: R.~Davies (davies@mpe.mpg.de), T.~Shimizu (shimizu@mpe.mpg.de), J.~Shangguan (shangguan@pku.edu.cn)}:
\and K.~Abd~El~Dayem\inst{2}
\and N.~Aimar\inst{10,13}
\and A.~Berdeu\inst{7,2}
\and J.-P.~Berger\inst{4}
\and G.~Bourdarot\inst{1}
\and P.~Bourget\inst{7}
\and W.~Brandner\inst{5}
\and Y.~Cao\inst{1}
\and C.~Correia\inst{10,13}
\and S.~Cuevas~Cardona\inst{16}
\and R.~Davies\inst{1}
\and D.~Defr\`ere\inst{9}
\and A.~Drescher\inst{1}
\and A.~Eckart\inst{6,14}
\and F.~Eisenhauer\inst{1,12}
\and M.~Fabricius\inst{1}
\and A.~Farah\inst{16}
\and H.~Feuchtgruber\inst{1}
\and N.M.~F\"orster~Schreiber\inst{1}
\and A.~Foschi\inst{2}
\and P.~Garcia\inst{10,13}
\and R.~Garcia~Lopez\inst{15}
\and R.~Genzel\inst{1,11}
\and S.~Gillessen\inst{1}
\and T.~Gomes\inst{10,13}
\and F.~Gont\'e\inst{7}
\and V.~Gopinath\inst{1}
\and J.~Graf\inst{1}
\and M.~Hartl\inst{1}
\and X.~Haubois\inst{26}
\and F.~Hau{\ss}mann\inst{1}
\and L.C.~Ho\inst{19,20}
\and S.~H\"onig\inst{8}
\and M.~Houll\'e\inst{4}
\and S.~Joharle\inst{1}
\and C.~Keiman\inst{16}
\and P.~Kervella\inst{2}
\and J.~Kolb\inst{7}
\and L.~Kreidberg\inst{5}
\and A.~Labdon\inst{26}
\and S.~Lacour\inst{2,7}
\and O.~Lai\inst{3}
\and S.~Lai\inst{23}
\and R.~Laugier\inst{9}
\and J.-B.~Le~Bouquin\inst{4}
\and J.~Leftley\inst{8}
\and R.~Li\inst{19,20}
\and B.~Lopez\inst{3}
\and D.~Lutz\inst{1}
\and F.~Mang\inst{1}
\and A.~M\'erand\inst{7}
\and F.~Millour\inst{3}
\and M.~Montarg\`es\inst{2}
\and N.~More\inst{1}
\and N.~Moruj\~{a}o\inst{10,13}
\and H.~Nowacki\inst{3}
\and M.~Nowak\inst{2}
\and S.~Oberti\inst{7} 
\and C.~Onken\inst{21,22}
\and J.~Osorno\inst{2}
\and T.~Ott\inst{1}
\and T.~Paumard\inst{2}
\and K.~Perraut\inst{4}
\and G.~Perrin\inst{2}
\and R.~Petrov\inst{3}
\and P.-O.~Petrucci\inst{4}
\and N.~Pourr\'e\inst{4}
\and S.~Rabien\inst{1}
\and C.~Rau\inst{1}
\and D.C.~Ribeiro\inst{1}
\and S.~Robbe-Dubois\inst{3}
\and M.~Sadun~Bordoni\inst{1}
\and M.~Salman\inst{9}
\and J.~Sanchez-Bermudez\inst{16}
\and D.~Santos\inst{1}
\and J.~Sauter\inst{5}
\and M.~Scialpi\inst{25}
\and J.~Scigliuto\inst{3}
\and J.~Shangguan\inst{19,20}
\and P.~Shchekaturov\inst{7}
\and T.~Shimizu\inst{1}
\and F.~Soulez\inst{17}
\and C.~Straubmeier\inst{6}
\and E.~Sturm\inst{1}
\and M.~Subroweit\inst{6}
\and C.~Sykes\inst{8}
\and L.J.~Tacconi\inst{1}
\and H.~\"Ubler\inst{1}
\and G.~Ulbricht\inst{18}
\and F.~Vincent\inst{2}
\and R.~Webster\inst{24}
\and E.~Wieprecht\inst{1}
\and J.~Woillez\inst{7}
\and C.~Wolf\inst{21,22}
}

\institute{
Max Planck Institute for extraterrestrial Physics, Giessenbachstra{\ss}e~1, 85748 Garching, Germany
% 2
\and LIRA, Observatoire de Paris, Universit\'e PSL, Sorbonne Universit\'e, Universit\'e Paris Cit\'e, CY Cergy Paris Universit\'e, CNRS, 92190 Meudon, France
% 3
\and Universite\'e C\^ote d’Azur, Observatoire de la C\^ote d’Azur, CNRS, Laboratoire Lagrange, Nice, France
% 4
\and Univ. Grenoble Alpes, CNRS, IPAG, 38000 Grenoble, France
% 5
\and Max Planck Institute for Astronomy, K\"onigstuhl 17, 69117 Heidelberg, Germany
% 6
\and $1^{\rm st}$ Institute of Physics, University of Cologne, Z\"ulpicher Stra{\ss}e 77, 50937 Cologne, Germany
% 7
\and European Southern Observatory, Karl-Schwarzschild-Stra{\ss}e 2, 85748 Garching, Germany
% 8
\and School of Physics \& Astronomy, University of Southampton, Southampton, SO17 1BJ, UK
% 9
\and Institute of Astronomy, KU Leuven, Celestijnenlaan 200D, 3001, Leuven, Belgium
% 10
\and Faculdade de Engenharia, Universidade do Porto, rua Dr. Roberto Frias, 4200-465 Porto, Portugal 
% 11
\and Departments of Physics and Astronomy, Le Conte Hall, University of California, Berkeley, CA 94720, USA
% 12
\and Department of Physics, Technical University Munich, James-Franck-Stra{\ss}e 1,  85748 Garching, Germany
% 13
\and CENTRA - Centro de Astrof\'{\i}sica e Gravita\c c\~ao, IST, Universidade de Lisboa, 1049-001 Lisboa, Portugal
% 14
\and Max Planck Institute for Radio Astronomy, Auf dem H\"ugel 69, 53121 Bonn, Germany
% 15
\and School of Physics, University College Dublin, Belfield, Dublin 4, Ireland
% 16
\and Universidad Nacional Aut\'onoma de M\'exico. Instituto de Astronom\'ia. A.P. 70-264, Ciudad de M\'exico, 04510, M\'exico
% 17
\and Univ. Lyon, Univ. Lyon 1, ENS de Lyon, CNRS, Centre de Recherche Astrophysique de Lyon UMR5574, 69230, Saint Genis-Laval, France
% 18
\and Dublin Institute for Advanced Studies, 31 Fitzwilliam Place, D02 XF86 Dublin, Ireland
% 19
\and Kavli Institute for Astronomy and Astrophysics, Peking University, Beijing 10087, People's Republic of China
% 20
\and Department of Astronomy, School of Physics, Peking University, Beijing 100871, People's Republic of China
% 21
\and Research School of Astronomy and Astrophysics, Australian National University, Weston Creek ACT 2611, Australia
% 22
\and Centre for Gravitational Astrophysics (CGA), Australian National University, Building 38 Science Road, Acton ACT 2601, Australia
% 23
\and Space \& Astronomy, Commonwealth Scientific and Industrial Research Organisation (CSIRO), Space \& Astronomy, P. O. Box 1130, Bentley, WA 6102, Australia
% 24
\and School of Physics, University of Melbourne, Parkville VIC 3010, Australia
% 25
\and INAF - Osservatorio Astrofisico di Arcetri, Largo E. Fermi, 50125 Firenze, Italy
%26
\and European Southern Observatory, Alonso de Cordova 3107, Vitacura, Casilla 19001, Santiago, Chile
}

   \date{Received Sep 2025}
 
  \abstract
{We present the first near-infrared interferometric data of a QSO at z=4. The K-band observations were performed with GRAVITY+ on the VLTI using all four UTs, detecting a differential phase signal that traces the spatially resolved kinematics for both the \hb\ and \hg\ lines in the broad line region. We fit the two lines simultaneously with an updated model that includes distinct rotating and conical outflowing components. For the best fit model, more than 80\% of the \hi\ line emission from the BLR originates in an outflow with a velocity up to $10^4$~km~s$^{-1}$. This is oriented so that our line of sight is along an edge of the conical structure, which produces the prominent blue wing on the line profile. A combination of anisotropic line emission and mid-plane opacity lead to the single-sided phase signal. The model is able to qualitatively match both the outflowing \civ\ line profile and the systemic \oi\ fluorescent emission. The derived black hole mass of $8\times10^8$~M$_\odot$ is the highest redshift black hole mass measurement to date obtained directly from BLR dynamics. It is an order of magnitude lower than that inferred from various single epoch scaling relations, and implies that the accretion is highly super-Eddington. With reference to recent simulations, the data suggest that this QSO is emitting close to its radiative limit in a regime where strong outflows are expected around a polar conical region.}


   \maketitle

\section{Introduction}
\label{sec:intro}
\nolinenumbers

The cosmic epoch at $3 \lesssim z \lesssim 5$ marks a key period between the end of reionisation at $z \gtrsim 6$ \citep{rob22,loe01} and the most active phase of galaxy formation at cosmic noon around $z \sim 2$ \citep{for20,tac20}.
There is an increasing interest in spatially resolving the molecular and ionised emission lines in galaxies on kiloparsec scales during this epoch \citep{fev20,wyl22,gal23,her25,ubl24}, because doing so can provide insights into the processes that have shaped the universe into what it is now.
The role of active galactic nuclei (AGN) is equally important because the black hole accretion rate is rising rapidly during this epoch until it peaks around the same time as galaxy formation \citep{air15,kim24}.
Indeed, both the major impact that feedback from AGN can have on galaxy growth \citep{fab12,kin15,fan23}, and the existence of local relations between black hole mass and their host galaxy properties \citep{fer00,geb00,gul09,mcc13,sag16}, indicate that supermassive black holes should co-evolve with galaxies \citep{kor13,hec14}.

Observations with JWST, as nicely illustrated by \citet{ada25}, have recently highlighted a population of AGN with luminosities overlapping those of local Seyfert nuclei, spanning the entire redshift range from cosmic noon to reionisation \citep{koc23,ubl23,gre24,mai24,mat24,suh25,juo25,sch25}.
The data have unexpectedly indicated that, compared to local relations, these black holes are overmassive with respect to their host galaxies.
The implication is that they must have grown very rapidly, requiring sustained near-Eddington, or even super-Eddington, accretion, depending on the mass of the seed black holes \citep{kok23,lar23,bog24,jeo24,mai24,meh24,suh25}.
However, concerns have been raised about the scale of uncertainties on both the black hole mass and host galaxy mass estimates, and on the use of scaling relations to derive black hole masses \citep{lam24,lup24,ada25,ber25}.

Complementary to these efforts are studies focussing on highly luminous QSOs, which are expected to drive the powerful outflows that have the greatest impact on host galaxy evolution \citep{bis17}.
Such QSOs are inferred to have black hole masses $M_\mathrm{BH} > 10^9$~M$_\odot$, with a significant fraction close to $10^{10}$~M$_\odot$, and so are radiating at around their Eddington luminosity \citep{vie18}.
The mass outflow rates derived from the prominent broad \oiii\ lines exceed 1000~M$_\odot$~yr$^{-1}$, and have kinetic energies around 1\% of the bolometric luminosity.
\cite{vie18} showed that they follow the expected trend \citep{sul11,sul15}, where those exhibiting prominent \oiii\ have moderately strong \civ\ emission with modest blue-shift, while QSOs without such strong \oiii\ emission have weaker \civ\ that is more strongly blue-shifted. 
These authors argued that this could be due to orientation effects affecting observation of the outflows in the broad line region (BLR); and comparison of the \civ\ blue-shift and FWHM led them to propose that the line profile is a combination of virialised and polar outflowing components.
Extremely luminous  QSOs at $z\sim3$ have also been targeted by mid-infrared JWST observations with the aim of resolving the host galaxy \citep{wyl22,vay24},
The data have revealed extended \oiii\ emission in the host galaxies, the outflows, and the circumgalactic medium.

While the properties of outflows on large scales correlate with AGN luminosity, there are also significant departures, for example due to the short timescale variability of AGN which can lead to a wide range of mass loading factors \citep{zub20}.
Similarly, how the outflow is driven, the relative importance of different gas phases, the role of dust, and the degree of clumpiness, can all impact the outflow properties and may change with distance from the AGN \citep{kin11,fau12,zub12,som15,war24}.
As such, resolving structures in these massive outflows nearer to their launching sites is central to understanding them, but doing so requires additional techniques due to the extremely small scales involved.
Velocity resolved reverberation mapping (RM) of nearby AGN has shown that there can be an outflow component to the BLR \citep{den09,gri13,du16a}.
More recently, near-infrared interferometry with GRAVITY has been used to spatially resolve the BLR \citep{gra18stu,gra20sha,gra21shi}, finding that in some local AGN, outflows can dominate the BLR kinematics \citep{gra24san}.
A QSO at $z \sim 2$ has now been observed with the same technique, resolving the BLR kinematics in a highly accreting QSO \citep{abu24}.
Observations of several more luminous QSOs at the same epoch have revealed clear signatures of radial flows in a significant fraction of their BLRs (GRAVITY+ Collaboration et al., in prep). 

In this paper we take the next step towards higher redshifts, and present interferometric data from GRAVITY+ that resolve the BLR in SMSS~J052915.80-435152.0 (hereafter J0529), a highly luminous QSO at z=3.96 \citep{wol24}.
Our focus is two-fold.
One aspect is to probe the AGN driven outflow on sub-parsec scales, where a broad-line region outflow may be launched.
The other is to make a dynamical estimate of the black hole mass, as part of a larger effort to understand whether the scaling relations derived locally can be reliably applied at high redshifts and if needed re-calibrate those relations.
Our observations are presented in Sec.~\ref{sec:obs}.
We provide a context for our analysis in Sec.~\ref{sec:lineprop} by reporting the BLR size and BH mass expected from scaling relations.
We then describe our BLR model in Sec.~\ref{sec:model} and present our fitting results in Sec.~\ref{sec:modresult}, speculating about why they differ so significantly from the scaling relations, and showing how the model can be used to match other emission lines.
We summarise our results in Sec.~\ref{sec:conc}.

\section{Observations and Data Reduction}
\label{sec:obs}
\nolinenumbers

The data analysed here were obtained with GRAVITY \citep{grav17} as part of programme 114.27UV.
They make use of the dual field mode of GRAVITY-wide \citep{grav22} and the new adaptive optics modules \citep{grav25}, which are upgrades towards GRAVITY+. As used here, GRAVITY combines the light of all four 8-m UTs of the VLT, to probe spatial scales of milli-arcsec and perform spectro-astrometry at 10~micro-arcsec levels.

The data were obtained on five nights from Sep. to Dec.~2024 with typical seeing (at zenith, and at 500~nm) on different nights of 0.5\arcsec\ and 0.7\arcsec, and coherence timescales of $\tau_0 = 5-10$~ms.
The MEDIUM-COMBINED dual-field wide mode was used, providing medium ($R \sim 500$) spectral resolution and combining the polarisations, and with more than 2\arcsec\ between the science target and fringe tracking reference star.
A total of almost 7.8~hr integration was obtained in 64 exposures as shown in Table~\ref{tab:obs}.
After processing the data with the GRAVITY instrument pipeline (v.~1.7.0b1), particularly noisy data sets were rejected, reducing the number used from 64 to 59.
An additional step was also made to compensate the wavelength dependent variability caused by the dispersion of air in the non-vacuum delay lines as described in Appendix~A of \citet{gra20sha}, using updated components calibrated from GRAVITY-wide observations.
Due to the limited signal-to-noise of individual science frames, it was applied on a night-by-night basis: the scaling derived from the combined frames on each night (excluding spectral regions near the \hb\ and \hg\ lines) was applied to each of those frames individually.
Although this approximation cannot take into account changes in delay line length due to tracking, it appears to be sufficient in these circumstances.
The resulting differential phase spectra from all nights were then combined with equal weighting, after rejecting (at most a few) deviant values in each spectral channel.

\begin{table}
\small
\caption{Observing log}
\label{tab:obs}
\centering
\begin{tabular}{ccccc}
\hline \hline
Date & seeing    & $\tau_0$ & No. exp. & NDIT$\times$DIT \\
     & (\arcsec) & (ms)     &          & (s)             \\
\hline
27 Sep 2024 & 0.47 & ~~8.4 &  3 & 4$\times$300 \\
14 Nov 2024 & 0.72 & ~~5.9 & 20 & 4$\times$100 \\
15 Nov 2024 & 0.48 &  10.4 & 22 & 4$\times$100 \\
12 Dec 2024 & 0.53 & ~~5.9 &  6 & 4$\times$100 \\
14 Dec 2024 & 0.72 & ~~4.7 & 13 & 4$\times$100 \\
\hline
\end{tabular}
\end{table}

An additional K-band spectrum was obtained with the ERIS adaptive optics imager and spectrometer \citep{dav23} as part of programme 112.25M3.
ERIS includes an integral field spectrometer (SPIFFIER) which was used here in seeing limited conditions (0.71\arcsec\ at zenith and 500~nm) with the largest 8\arcsec\ field of view, and 
the K-band grating covering 1.96-2.45~$\mu$m at a resolution of $R\sim5000$.
The data were taken as 11 dithered exposures of 300~s, yielding a total integration time of 55~mins, and reduced with the ERIS instrument pipeline (v~1.6.4).
Verification of the wavelength calibration was performed using OH sky emission lines.
Telluric correction and flux calibration were performed using the B9\,IV standard star HD~38343 which has K=7.87~mag.
Following this, additional steps were performed to interpolate over the noisiest pixels in the datacube and reduce wavelength dependent variations in the background level.
The flux density of the QSO was measured from a spectrum extracted in a large 1.25\arcsec\ aperture, yielding K = 13.88~mag.

\begin{figure}[ht]
\centering
\includegraphics[width=0.70\hsize]{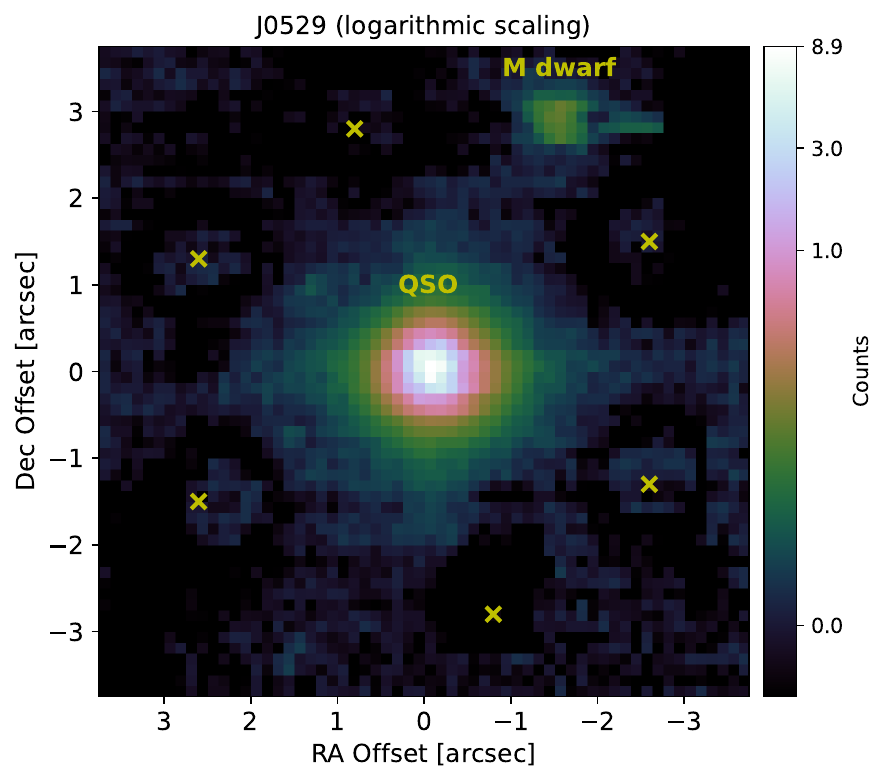}
\caption{K-band image of J0529 from ERIS. The 6 yellow crosses indicate regions where the sky background is affected, at a very faint level, by the in-frame dithering. The QSO shows no structure indicative of strong lensing, and the only other source in the field is the M dwarf star about 3\arcsec\ away and 4~mag fainter.}
\label{fig:eris_img}
\end{figure}

\subsection{Image}
\label{sec:image}

An image was extracted from the datacube by summing channels in the range 2.028--2.290~$\mu$m, indicating that the spatial resolution achieved on the QSO itself in the K-band was 0.43\arcsec.
The image is shown in Fig.~\ref{fig:eris_img} on a logarithmic scaling to highlight faint sources.
The QSO shows no sub-structure, and the only other source in the field is the M dwarf star previously noted by \cite{wol24} a few arcsec north of the QSO.
This strengthens the assessment made by those authors that there is no sign of strong lensing, and hence the luminosity is intrinsic.

\begin{figure*}[ht]
\centering
\includegraphics[width=0.8\hsize]{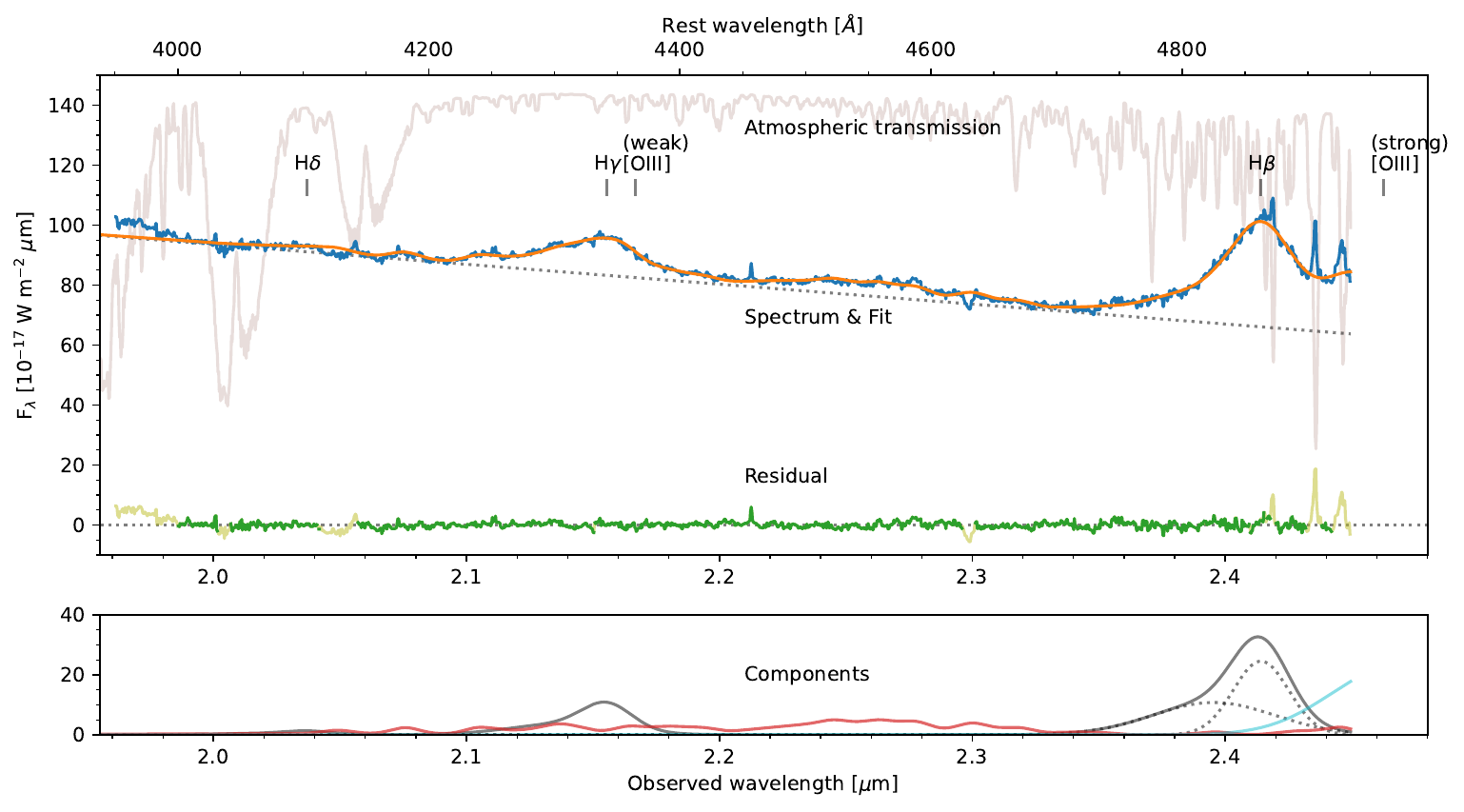}
\caption{K-band spectrum of J0529 from ERIS. The upper panel shows the data (blue; the narrow features at the long end of the band are due to imperfect atmospheric correction) and the fit (orange), with the \hi\ lines marked (the location of the \oiii\ lines are also marked; the 4363~\AA\ line is expected to be 10-100 times weaker than those at 4959~\AA\ and 5007~\AA, the peaks of which are out of the spectral range). Also shown is the atmospheric transmission and the residual, indicating which regions have been excluded from the fit.
The lower panel shows the components of the fit: the \hi\ lines (black), \feii\ complex (red), and a component that represents either poor slope correction at the band edge (as seen also at the short end of the band) or a possible blue wing on the \oiii\ line.}
\label{fig:erisspec}
\end{figure*}

\subsection{Spectral decomposition}
\label{sec:spec}

A final spectrum with higher signal-to-noise (S/N) of $\sim$85 (but slightly lower flux) than that above was extracted in a 0.62\arcsec\ aperture better matched to the FWHM.
This is used to recover the spectral profile of the broad \hi\ lines, which requires decomposition of the spectrum in order to deblend them from the \feii\ features which are known to be strong in highly accreting QSOs \citep{bor92,du16b}.
Since it is already clear from Fig.~\ref{fig:erisspec} that the \hi\ lines have a prominent blue wing, we fit these with a pair of Gaussians.
This is done solely to match the overall profile and we do not consider the Gaussians individually.
We use the same velocity profile for both \hi\ lines, which, although not physically required, is necessary due to limitations of the data.
In addition we required zero velocity offset between the \hi\ lines, which was an important constraint on the fit.
The reason is that the blue wing of the \hg\ line is strongly blended with \feii\ and so is primarily constrained by the \hb\ line. 
On the other hand the red side of the \hg\ line does provide a primary constraint, whereas the \hb\ line is affected by both atmospheric absorption (leading to narrow features and also affecting the slope) and potentially some emission from a blue-shifted wing of \oiii\ line emission, the peak of which would be just out the spectral range.
While we do not have direct evidence for \oiii\ (and indeed the \oiii\ line at 4363~\AA\ is not detected although it is expected to be a factor 10-100 weaker), in luminous high-z samples it can be associated with an outflow as indicated by the strong blue-shifted \civ\ emission described in Sec.~\ref{sec:linecomp} \citep{mar17}.
The intrinsic continuum is expected to be a power law, but over this narrow wavelength range we have adopted a linear function.
We consider several templates for the optical \feii\ emission, including those based on I~Zw~1 \citep{ver04}, Mrk~493 \citep{par22}, a composite based on several QSO spectra \citep{tsu06}, and an adaptable template derived from multiple AGN \citep{kov10}.
After trying these options, we adopted the template from \citet{par22} because it provides the best match to the \feii\ emission on the blue side of the \hg\ line; in other regions the templates were equivalently good.
Although the properties of the \feii\ complex were not tied to the Balmer lines, the broadening of $\sim1500$~km~s$^{-1}$ FWHM is comparable to the width of the Balmer lines as one would expect if they arise in similar locations \citep{hu15}.
Finally, specific features in the spectrum that were clearly due to atmospheric effects were excluded during the fitting process.
The resulting fit has $\chi^2_\mathrm{red}= 0.96$.
The observed spectrum and its model fit, showing the contribution of the \feii\ template and \hi\ line profiles, are shown in Fig.~\ref{fig:erisspec}.

\section{Emission lines: inferences and comparisons}
\label{sec:lineprop}
\nolinenumbers

From the spectral fit described above, we calculate a line luminosity\footnote{Adopting $H_0=70$~km~s$^{-1}$~Mpc$^{-1}$, $\Omega_\mathrm{M}=0.286$, and $\Omega_\mathrm{vac}=0.714$; and using the calculation provided by \cite{wri06}.} of $L_\mathrm{H\beta} = 2.0\times10^{45}$~erg~s$^{-1}$, and can extrapolate the optical continuum luminosity to be $\lambda L_\lambda = 2.3\times10^{47}$~erg~s$^{-1}$ at 5100~\AA.
The equivalent width of \hb\ is $\sim200$~\AA\ in the observed frame and hence $\sim40$~\AA\ in the rest frame, putting it within the typical range of $20-80$~\AA\ found for most QSOs \citep{ost78,hu08}.

The line ratio \hg/\hb\ $ = 0.33$ is lower than the value of 0.47 for case B recombination at $10^4$~K \citep{ost89}. It is also marginally lower than the range of values $0.45\pm0.08$ reported by \cite{mur07} for broad line AGN, although it is consistent with the lowest values these authors found for several of their sample.
We note that H$\delta$ is within our spectral range, and we find\footnote{the S/N of the line ratios is relatively high, in this case $\sim8$, because their shapes are tied to be the same.} $H\delta$/H$\beta = 0.04$ which is slightly less than the lowest values reported by the same authors.
As such the line ratios appear physically plausible, and indicate that there might be some modest extinction towards the broad lines.

The full line profile has FWHM$ = 3907$~km~s$^{-1}$ and $\sigma = 2820$~km~s$^{-1}$ so that FWHM/$\sigma = 1.39$.
This ratio, has been discussed significantly in the literature \citep{col06} and is known to vary between AGN.
The value we find is significantly lower than the canonical value of 2 often adopted for local AGN when deriving scaling relations \citep{col06,woo15}, and is a direct result of the blue wing on the line profile.

We also calculate the ratio\footnote{An alternative that is also commonly used is the ratio of the equivalent widths \citep{bor92,she14}. In practice it makes little difference because the continuum flux density changes very little between the corresponding wavelengths.} 
$\rfe = F_\mathrm{Fe}/F_\mathrm{H\beta}$ where $F_\mathrm{Fe}$ is the flux of \feii\ in the range 4434-4684~\AA\ \citep{du19}.
As noted above, this is indicative of the accretion rate $\dot{M}_\mathrm{acc}$ and, in a sample of AGN selected to be super-Eddington, \cite{du19} find many of their objects have $\rfe \gtrsim 1$. 
Despite the high luminosity of J0529, our value of $\rfe = 0.32$ is modest compared to that sample.

The estimates above for the optical continuum luminosity and broad line luminosities and widths, enable us to make use of scaling relations to estimate both $R_\mathrm{BLR}$ and $M_\mathrm{BH}$.
These single epoch estimates provide a context for the dynamical modelling that we perform on the GRAVITY+ data in Sec.~\ref{sec:modresult}.

\subsection{BLR size from scaling relations}

\begin{table}
\small
\caption{BLR size estimates from \hb\ scaling relations}
\label{tab:blrsize}
\centering
\begin{tabular}{lll}
\hline\hline
$R_\mathrm{BLR}\ (pc) $ & Reference & Note \\
\hline
1.74 & \cite{ben13} & commonly used \\
1.47 & \cite{du18} & lower $\dot{M}_\mathrm{acc}$ subsample \\
1.22 & \cite{du19} & without $\mathcal{R}_\mathrm{Fe}$ correction \\
0.94 & \cite{du19} & with $\mathcal{R}_\mathrm{Fe}$ correction \\
0.65 & \cite{woo24} & 47 AGN sample \\
0.50 & \cite{du18} & higher $\dot{M}_\mathrm{acc}$ subsample \\
0.48 & \cite{woo24} & 240 AGN sample \\
\hline
\end{tabular}
\end{table}

We have derived the BLR size $R_\mathrm{BLR}$ from 4 different relations and summarised them in Table~\ref{tab:blrsize}.
One of the best known $R_\mathrm{BLR}-L_\mathrm{AGN}$ relations between the AGN luminosity and size of the BLR is that of \cite{ben13}.
The AGN luminosities were derived using a surface brightness decomposition to correct for the host galaxy; and the BLR sizes were derived from a simple measure of the time lag obtained from RM.
An emphasis to AGN with high accretion rates was given by \cite{du18}, who derived different relations depending on the accretion rate. We show here the relation for both lower and higher accretion rates, which yield very different results.
We also consider a version of that relation which was updated by \cite{du19} to include an explicit correction for the accretion rate via the $\mathcal{R}_\mathrm{Fe}$ parameter.
For J0529 this provides a substantial reduction to the inferred BLR size compared to \citet{ben13}, although for J0529 the $\rfe$ correction itself is only a 25\% effect.
Finally, we consider estimates from \cite{woo24} who provided two fits.
We include both, one based on a combination of 47 AGN from two specific samples, and one derived from 240 AGN that are mostly from different sources in the literature.
These authors found no strong $\mathcal{R}_\mathrm{Fe}$ dependent deviation from a $R_\mathrm{BLR}-L_\mathrm{AGN}$ relation, but they did note that the slope of the relation is likely to be shallower than that popularly used.
These differences highlight the uncertainties associated with employing scaling relations. 
Since there is also a scatter (typically reported as $\sim0.2$~dex) associated with each relation, here we can conclude only that $R_\mathrm{BLR}$ is likely in the range 0.5-1.7~pc.

\begin{table*}
\caption{BH mass estimates from \hb\ scaling relations}
\label{tab:bhmass}
\centering
\begin{tabular}{lll}
\hline\hline
$\log{M_\mathrm{BH}/M_\odot}$ & Reference & Note \\
\hline
10.43 & \cite{dal20} & Eq.~38 using $\sigma_\mathrm{H\beta}$ \& L$_\mathrm{H\beta}$ \\
10.24 & \cite{wol24} & continuum modelling; also \civ\ \& \mgii \\
10.16 & \cite{woo15} & Eq.~2 using $\sigma_\mathrm{H\beta}$ \& L$_{5100}$ \\
~~9.87 & \cite{rei13} & Eqs.~2, 3, \& 5; converting FWHM$_\mathrm{H\beta}$ \& L$_{5100}$ to H$\alpha$\\
~~9.58 & \cite{vie20,bon14} & Eq.~4 or Eq.~2 respectively, using FWHM$_\mathrm{H\beta}$ \& L$_{5100}$ \\
\hline
\end{tabular}
\end{table*}

\subsection{Black hole mass from scaling relations}

In a similar approach, we estimate the black hole mass $M_\mathrm{BH}$ from several different relations which are summarised in Table~\ref{tab:bhmass}.
The mass reported by \cite{wol24} in their discovery paper was $\log{M_\mathrm{BH}/M_\odot}=10.24$ based on both modelling of the continuum shape and use of the \civ\,$\lambda$1549 and \mgii\,$\lambda$2799 line widths.
A recent re-assessment of the use of \hb\ to estimate $M_\mathrm{BH}$ was provided in Eq.~38 of \cite{dal20}, who included a (statistical) correction for Eddington ratio $\lambda_\mathrm{Edd}$. 
These authors also argued that the dispersion $\sigma$ is a less biassed tracer of the line width than FWHM (although use of $\sigma$ for profiles with Lorentzian-like wings also poses difficulties, \citealt{san25}); and prefer to use the line rather than continuum luminosity in order to avoid bias from core dominated radio sources.
An important part of the relation is converting the virial mass into a black hole mass via the $f$ parameter, and \cite{dal20} adopt $\log{f}=0.683$ from \cite{bat17}.
For J0529, the inferred BH mass is $\log{M_\mathrm{BH}/M_\odot}=10.43$, which is high probably because of the bias on both the line dispersion and luminosity caused by the blue wing.
An earlier relation which did not attempt to make any corrections, and was based on the $R_\mathrm{BLR}-L_\mathrm{AGN}$ relation of \cite{ben13}, is given as Eq.~2 of \cite{woo15}, for which they adopt $\log{f}=0.05$ (a very different value than used by \cite{dal20}, emphasizing that it depends on how the scaling relation is set up and whether FWHM or $\sigma$ is used).
From this relation we infer a value of $\log{M_\mathrm{BH}/M_\odot}=10.16$ using the full line profile.
\cite{rei13} derived a relation based on the continuum luminosity and FWHM of the \hb\ line, providing conversions for these quantities to H$\alpha$ luminosity and FWHM in Eqs.~2 and~3.
This also builds on the $R_\mathrm{BLR}-L_\mathrm{AGN}$ relation of \cite{ben13}, and uses the conversions to H$\alpha$ from \cite{gre05}.
We can put our measured values into those equations, apply the conversions, and hence estimate a mass from their Eq.~5.
Doing this, and noting that they use $\log{f}=0$, yields $\log{M_\mathrm{BH}/M_\odot}=9.87$.
The final relation we consider is Eq.~4 from \cite{vie20} which was applied to luminous QSOs at $z\sim2$, and which was taken directly from Eq.~2 in \cite{bon14} where it was applied to X-ray obscured red QSOs at $1.2<z<2.6$.
As for the previous relation it uses FWHM$_\mathrm{H\beta}$ and L$_{5100}$, yielding $\log{M_\mathrm{BH}/M_\odot}=9.58$.
The conclusion here is that $M_\mathrm{BH}$ is associated with even more uncertainty than $R_\mathrm{BLR}$, and that there may be a systematic difference when using $\sigma$ vs FWHM.
For J0529 the scaling relations tend to give values around $\log{M_\mathrm{BH}/M_\odot}\sim10$.

\section{Description of the BLR model}
\label{sec:model}
\nolinenumbers

We make use of the phenomenological model described by \cite{pan14}, our implementation of which is given in detail in \cite{gra20sha} with summaries in \cite{gra21shi} and \cite{gra24san}.
The model implements the BLR as an ensemble of clouds moving in the potential of the central black hole.
Physically, clouds in the BLR are likely to be transient objects, plausibly formed by shock compression due to gas collisions on small scales (as has been adopted to model the ISM on scales of $\sim1$-10~pc in AGN, \citealt{vol18}).
However, irrespective of the true nature of the clouds, in our phenomenological model the use of the term is for convenience. 
The clouds should be taken to represent line emitting test particles rather than physical entities, because the model parametrises the distribution and kinematics of the line emission rather than the entire gas content of the BLR.

The model of \cite{pan14} is a good representation of the physics motivating the FRADO (Failed Radiatively Accelerated Disk Outflow) model, proposed by \cite{cze11} and developed by \cite{cze17} and \cite{bas18}.
In this concept, the anisotropic emission from the accretion disk and the effects of shielding enable dust to exist in a thin disk at radii smaller than the nominal sublimation radius.
The radiation pressure from the hot disk can accelerate a dusty gas cloud vertically until it has a direct line of sight to the accretion disk.
At this point the dust sublimates, and the cloud continues ballistically before falling back down to the disk.
The BLR therefore forms a flared region above the disk plane, as proposed by other authors \citep{goa12,ram18,hop25b}.
These models are consistent with the dust having a bowl-shaped form that was proposed to explain the difference in dust sizes between interferometric and reverberation mapping measurements \citep{gra24cao}.

A limitation of the FRADO model is that it only considers vertical motion of the clumps, and does not address their radial motion.
Once clouds are driven high enough that the dust sublimates, the direct line of the sight they have to the AGN is also expected to lead to an outward acceleration.
This may be either by UV-line driving, as is believed to be happening for ultra fast outflows (UFOs) from the accretion disk \citep{miz21}, or as a magneto-centrifugal wind \citep{emm92}, which has also been proposed as an outflow mechanism on 1-10~pc scales \citep{vol18}.
The BLR outflow will form part of a hierarchy of outflows, located between the small scale UFOs and the radiatively driven outflow that accelerates dusty gas to create geometrically thick obscuration \citep{ram17,cha17,hoe19}.
Physically, and based on simulations \citep{kud23,kud24,hop25a,hop25b}, one would expect outflow to be driven at the surface of the BLR and inflow to occur closer to the disk plane.
However, in the \cite{pan14} model, radial motions are included in a way that allows a fraction of the clouds throughout the bulk of the BLR to be either all inflowing or all outflowing (and unbound) on trajectories that are essentially radial.

\begin{figure}[ht]
\centering
\includegraphics[width=0.9\hsize]{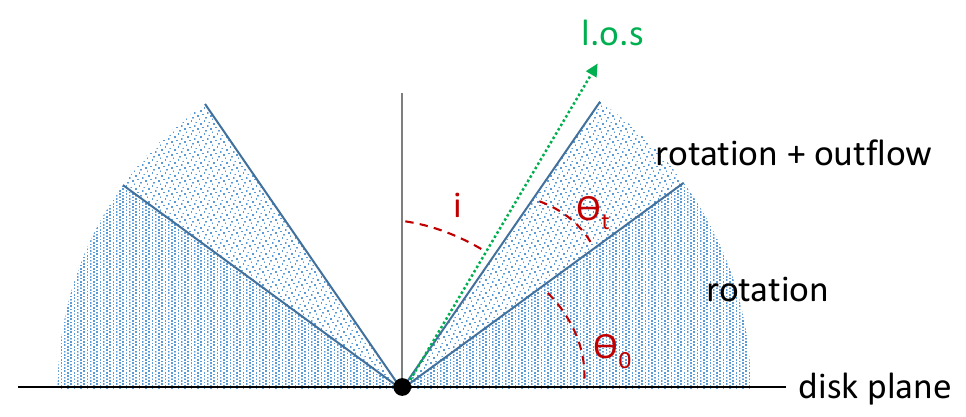}
\caption{Sketch of modified BLR model. Only the half of the BLR above the disk plane is shown; it is symmetric below, with the emission potentially modified by anisotropy and mid-plane opacity. The lower part with a flare angle $\theta_0$ corresponds to a rotating disk; the upper part with a flare angle $\theta_t$ can be outflowing.}
\label{fig:newmodel}
\end{figure}

In order to better physically motivate the way radial flows are handled in our phenomenological model, we have updated its implementation while keeping the model as simple as possible.
The original flared disk model is used to represent the static part of the BLR described by \cite{cze17} and \cite{bas18}.
It is unchanged except that outflows are excluded, while inflow is in principle allowed.
However, after initial iterations of the model, it was clear that for J0529 the fraction of inflowing clouds was small enough that, for simplicity, it could be excluded entirely with negligible impact.
Thus, in this case, the motions in the disky part of the model are purely rotational.

\begin{figure*}
\sidecaption
\includegraphics[width=12cm]{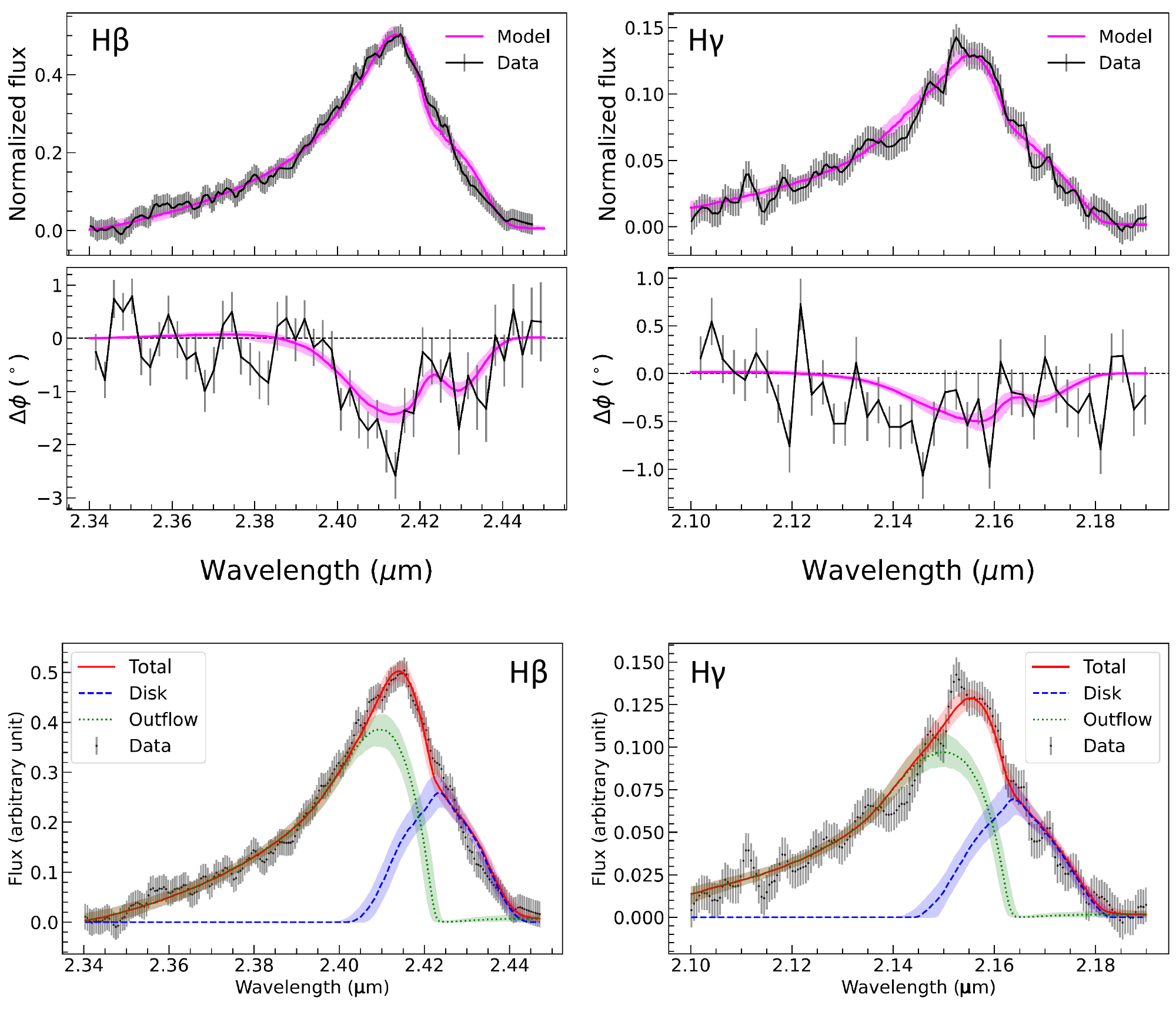}
\caption{Flux spectrum, averaged differential phase spectra, and line decomposition. 
The top row shows the continuum subtracted spectrum of the two emission lines, and the best fit model. 
For each of the \hb\ (left) and \hg\ (right) lines, the middle row shows the average phase spectra  which are created from the three baselines with the strongest signal (this is for visualisation purposes only; phase spectra for the individual baselines are shown in Fig.~\ref{fig:allphases}).
The bottom row shows the decomposition of the fitted line profile (red) into the outflow (green) and disky (blue) components. The outflow dominates the total line flux, including much of the apparent core of the line profile.}
\label{fig:avgphase}
\end{figure*}

A specific outflow component is added as an additional wedge that extends above the upper surface of that flared disk as shown in Fig.~\ref{fig:newmodel}.
Again to avoid including more free parameters, we assume that the outflow and the disk components are spatially connected.
It represents the concept put forward by \cite{elv00} as well as the outflowing part in the simulations of \cite{kud23,kud24} and \cite{hop25a}, and has been modelled explicitly by \cite{yon17,yon20}.
The total angular height of the BLR is then the sum of $\theta_0$ for the flared disk and $\theta_t$ for the outflow.
Thus a situation where $\theta_0 + \theta_t = 90^\circ$ would mean that there are clouds on orbits perpendicular to the disk plane.

Clouds in the outflow are required to rotate around the central black hole with the same circular velocity they would have if they were in the midplane at the same projected radius. 
They are also outflowing with a radial velocity $V_\mathrm{out}(r)$.
The original model defined the outflow velocity at a radius $r$ to be the same as the escape velocity 
$V_\mathrm{esc} = \sqrt{2 G M_\mathrm{BH}/r}$ at that radius.
A couple of examples show that this was a reasonable assumption.
For $R_\mathrm{BLR}$ and $M_\mathrm{BH}$ derived in NGC~3783 \citep{gra21shi}, where about half the clouds are on unbound orbits, $V_\mathrm{esc} = 5700$~km~s$^{-1}$ at the average radius of the BLR is only a factor $2-3$ greater more than the 1800~km~s$^{-1}$ FWHM and within the full line profile;
Similarly, for the nominal size (1~pc) and mass ($10^{10}$~M$_\odot$) inferred in Sec.~\ref{sec:lineprop} for J0529, V$_\mathrm{esc} = 9400$~km~s$^{-1}$ is only a factor of a few greater than the dispersion of the line profile including the blue wing.
However, the assumption that $V_\mathrm{out} = V_\mathrm{esc}$ is also a limitation, and
can lead to unphysical geometries for prominent fast outflows for which the model is not suited.
Rather than follow the same prescription for $V_\mathrm{out}$ we apply an improvement suggested by \cite{pan14} to link it to the escape velocity at the wind launching radius \citep{mur95,lon23}.
Our implementation ties V$_\mathrm{out}$ more flexibly to the escape velocity at the minimum radius and still enables it to decrease with radius:
\[
V_\mathrm{out}(r) = f_V \ \sqrt{\frac{2 G M_\mathrm{BH}}{R_\mathrm{min}}} \ \left(\frac{r}{R_\mathrm{min}}\right)^{\alpha_V}
\]
where $R_\mathrm{min}$ is the minimum radius for the cloud distribution, $f_V$ is a scaling factor (in the range one to a few) for the escape velocity at $R_\mathrm{min}$, and $\alpha_V$ is an index in the range $-1 \le \alpha_V \le 0$ allowing $V_\mathrm{out}$ to decrease with increasing radius.
We argue that such a radial decrease should be expected because reverberation mapping of the \civ\ line, which has much a higher excitation potential than \hi, tends to yield a smaller radial scale than \hb\ \citep{lir18,she24}; and the line is often blue-shifted with respect to the \hi\ lines \citep{coa17,tem23}, as also illustrated for J0529 in Fig.~\ref{fig:linecomp}.
Thus it is reasonable to conclude that the outflow velocity is higher at smaller radii.

Finally, and again to keep the number of free parameters as low as possible, 
the vertical angular distribution of clouds in each of the outflowing and rotating regions is fixed to be uniform (rather than, as done in the original model, using $\gamma$ to control whether the clouds are concentrated more towards the top edge of the distribution).
However, because there is a new parameter $f_\mathrm{out}$ defining the fraction of clouds in the outflow, it is possible for the cloud density to differ between the outflowing versus rotating regions.

\section{The fitted BLR model}
\label{sec:modresult}
\nolinenumbers

We fit the \hb\ and \hg\ lines simultaneously in a similar way to that described in \citet{kuh24}: all the parameters in the fit are tied to be the same for both lines except for the two that define their radial distribution. 
These are the $\beta$ parameter that defines the form of the distribution, and the scale factor $F$ which is linked to the average radius $\mu$ such that $F = R_\mathrm{min}/\mu$, where $R_\mathrm{min}$ is the same minimum radius for both lines.
The details of the fit are given in Appendix~\ref{app:details}, including Table~\ref{tab:model} summarising the best fitting parameters corresponding to the model, and Fig.~\ref{fig:allphases} of the resulting spectra and differential phases for each baseline.
As a complement to this, the top two rows of Fig.~\ref{fig:avgphase} show the data and fitted model for each of the line profiles as well as the sum of the phases from baselines UT~4-2, 4-1, and 3-2 which contain the strongest signal.
The three main points from the phase data in Fig.~\ref{fig:avgphase} are (i) the phase signal for \hg\ is about 1/3 the strength of that for the \hb\ line, as expected from their equivalent widths; (ii) the phase signal is single-sided which is indicative of predominant outflow rather than the S-shape associated with rotation; and (iii) there are two distinct peaks in the phase signal, which are related to the separate outflowing and disky components.
This last point is developed further in the bottom panels of Fig.~\ref{fig:avgphase} which show the contribution of the outflowing and disky components to the total line profile.
It is clear that emission from the outflow is not just limited to the blue wing of the profile, but makes up a substantial part of the core as well. 
In contrast the disky component has only a subdominant contribution more towards the red side of the line.
Histograms of the posterior distributions for all the parameters are presented in Fig.~\ref{fig:histograms}, and the most relevant results for our analysis are additionally shown in the corner plot in Fig.~\ref{fig:corner}.
These figures indicate that the fit is well constrained; and that while other potential solutions might exist, these are at a much lower probability.

\begin{figure}[ht]
\centering
\includegraphics[width=1.0\hsize]{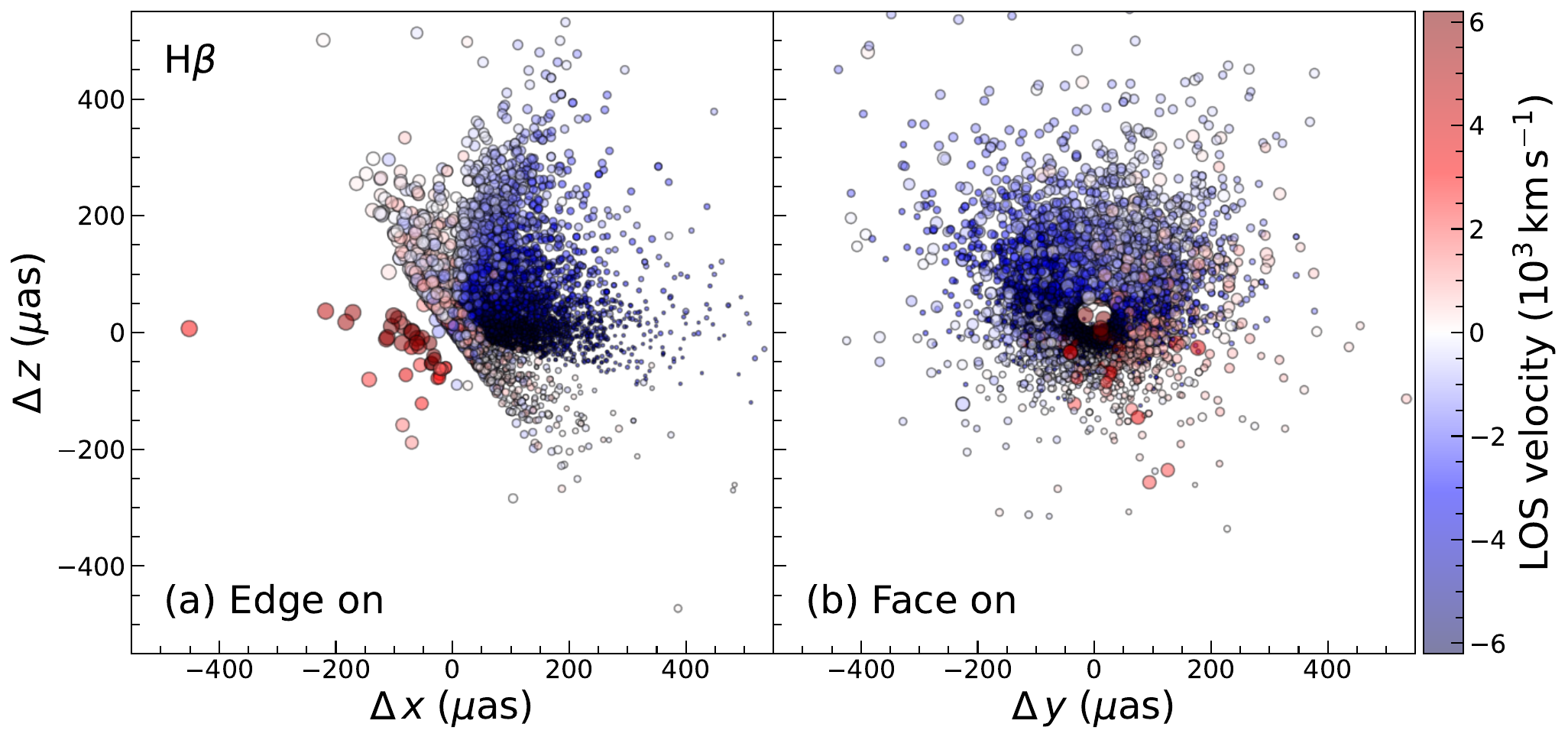}
\caption{Views of the BLR based on the model. Left: edge-on view, where our line of sight is from the right. Right: face-on view, as we would see the BLR. The clouds are coloured according to their line-of-sight velocity, and their size corresponds to their apparent brightness (invisible clouds are not shown). The edge-on view shows the impact of the high mid-plane opacity (few clouds on the far side of the BLR) and anisotropy (clouds for which we see the illuminated side are brighter).}
\label{fig:blrviews}
\end{figure}

\subsection{Systemic velocity}
\label{sec:sysvel}

Our spectrum and analysis imply a modest update of the redshift of J0529 with respect to the value of $z=3.962$ reported by \citet{wol24}.
Taking the peak of the \hi\ lines as representative of the systemic velocity, we find $z=3.965$.
However, the model indicates that the systemic velocity is shifted with respect to the line peak by about 1200~km~s$^{-1}$.
This shift is denoted by the parameter $\epsilon=0.0046$ in Table~\ref{tab:model}, which is the relative shift of the zero (systemic) velocity wavelength $\lambda_0$ from the initial value via the relation $\lambda_0 = (1+\epsilon) \lambda_\mathrm{initial}$ -- hence the value of $\epsilon$ is less important than that of $\lambda_0$ which it generates. 
The posterior distributions in Fig.~\ref{fig:histograms} show that $\epsilon$, and therefore also $\lambda_0$, is tightly constrained by the model.
While there is no specific feature in the data that determines what $\lambda_0$ must be, it arises because of the way the model is defined and how it is constrained by the data.
The lower part of the conical outflow is aligned towards us, giving rise to the high velocities in the long blue-shifted tail where the line flux is lower because we are seeing the back side of the clouds, and there is only a weak phase signal because there is no spatial offset. 
In contrast, Fig.~\ref{fig:blrviews} shows that the upper part of the same conical outflow is viewed at a more inclined angle.
Here, the lower line-of-sight velocities contribute more to the central part of the line profile as shown in Fig.~\ref{fig:avgphase}, and the emission is stronger because we are seeing more the illuminated side of the clouds.
Importantly, the spatial offset of this part of the outflow is what leads to the strong one-sided phase signal near the line centre in Fig.~\ref{fig:avgphase}.
Because of these constraints, the symmetric disk contribution must be slightly redshifted with respect to the line peak.
As a test to verify this outcome, we tried modelling the data with a prior on $\epsilon$ forcing the systemic velocity to be close to the line peak. 
There were two peaks in the posterior distribution.
The prior gave more weight to the solution close to the line peak, but this model is physically less plausible due to the high inclination and even lower black hole mass.
The higher peak was nearer the edge of the prior range and had a redshifted systemic velocity, which tends to confirm the original result.
This test emphasizes that the uncertainty on the systemic velocity results from matching the data -- which provide strong constraints on possible geometries -- within the scope of the model.
Taking this uncertainty into account, the resulting redshift is $z=3.984 \pm 0.002$.
We note, however, that the difference to the previous estimates is small enough that the impact on the inferred properties of the QSO (luminosity, black hole mass, etc) is negligible.

\subsection{BLR geometry and size}
\label{sec:blrsize}

The fitted parameters in Table~\ref{tab:model} show that $f_\mathrm{out} = 83$\% of the clouds are in the outflow, while the remaining 17\% are in the disky component on bound elliptical orbits.
This disparity could indicate that the outflow dominates the BLR. 
But an alternative interpretation of this effect could be that the outflow is obscuring the disky part of the BLR. The model cannot distinguish between these options: it fits the properties of the observable line emission but does not address the physical processes that generate it.

Our line of sight is near the top edge of the outflow, because the inclination is $i = 33^\circ$ from face-on while the total angle from the mid-plane subtended by BLR clouds is $\theta_0 + \theta_t = 60^\circ$.
If the line of sight were oriented further into the outflowing wedge, one might expect to see broad absorption lines in the rest-frame optical spectrum \citep{elv00} but these are not observed \citep{wol24}.
The emission is anisotropic with $\kappa = -0.45$ indicating that the side of the illuminated side of each cloud emits preferentially.
Equally important is the mid-plane opacity with $\xi=0.016$ meaning that the emission from clouds behind the mid-plane is almost entirely blocked, consistent with the basis of the FRADO model in which a dusty disk extends to small radii at low scale heights, and with the optically thick nature of super-Eddington accretion disks.
This leads to the configuration in Fig.~\ref{fig:blrviews}, where the clouds are coloured according to their line-of-sight velocity and have a size corresponding to their apparent brightness.

The average radius for the BLR is $\mu = 1.0$~pc for \hb\ and $\mu = 0.9$~pc for \hg.
While these are the same within their uncertainties (see Table~\ref{tab:model} and Fig.~\ref{fig:histograms}), this difference is expected both observationally, based on RM delays for different lines \citep{ben10}, and on modelling of radiation pressure confined BLR clouds \citep{net20}.

While the size appears to match that inferred by applying a $\rfe$ correction to the R-L relation favoured by \cite{du19}, this does not provide an explanation of why $\rfe$ is so low.
Indeed, from the plot of FWHM versus $\rfe$ in \cite{she14}, both of these values for J0529 seem rather typical of QSOs generally.
Instead we propose that much of the deviation from the scaling relations that has previously been attributed to luminosity or accretion rate, could, analogous to the \civ\ line, be due to the impact of the outflow on the \hi\ line properties.
Indeed, the disky part of the BLR in Fig.~\ref{fig:avgphase} contains only 27\% of the total line luminosity.
Applying this to the typical equation which has a typical form
$R_\mathrm{BLR} \propto 0.5 \log{L}$
where $L$ is usually taken to represent the total AGN luminosity.
If instead we consider only the luminosity associated with the disky part of the BLR tracing rotationally supported gas (using the line luminosity as a suitable proxy), then the inferred size is reduced by a factor $\sim2$.
This would bring the estimate from \cite{ben13} into good agreement with our measurement.

Similarly, we would find $\rfe = 1.44$ which is what one would expect for a highly accreting QSO, particularly in the context of Eddington ratio discussed below.
However, applying this new value would lead to a revised BLR size from \cite{du19} that is much too small.
We therefore speculate that the over-estimate of the BLR size that has been discussed in the literature for highly accreting and highly luminous QSOs could be related to the impact of BLR outflows on the line properties.

\subsection{Dynamical black hole mass}
\label{sec:bhmass}

The dynamical black hole mass we derive is $M_\mathrm{BH} = 8\times10^8$~M$_\odot$, which is lower than those inferred from scaling relations by an order of magnitude.
The low mass is largely driven by the small rotational velocity required in the model, which has to be less than the line width in order to allow for the outflow.

The form of the scaling relation used to estimate black hole mass is typically 
$M_\mathrm{BH} \propto 0.5 \log{L} + 2 \log{\sigma}$
where $L$ represents AGN luminosity and $\sigma$ (or FWHM) the line width.
If, as above, we take $L$ to be a proxy for the line luminosity (noting that line luminosity is used explicitly in some relations), applying this to only the disky part of the BLR profile which has $\sigma=1040$~km~$^{-1}$ and contains 27\% of the luminosity, reduces the inferred $M_\mathrm{BH}$ by a factor 14 (or a factor 5 if using FWHM, due to the differing FWHM/$\sigma$ ratios).
This use of the rotational velocity of the BLR rather than the full line width (which is broadened by the outflow) would bring all the single epoch estimates much closer to the BH mass we have measured, and supports the speculation above that the outflow could be a major contributor to the bias in single epoch estimates for luminous QSOs.

\subsection{Eddington ratio}
\label{sec:lambda_edd}

The Eddington ratio $\lambda_\mathrm{Edd}$ is determined by the black-hole mass and the bolometric luminosity $L_\mathrm{bol}$, which is commonly derived from a monochromatic luminosity and a bolometric correction $k_\mathrm{bol}$. Also, the emission from the accretion disc requires an anisotropy correction that depends on the viewing angle, and to a smaller extent on the black-hole spin since the deflection of light due to gravitational lensing is stronger for disc annuli that are closer to the black hole \citep{nem10,lai23}. Generic $k_\mathrm{bol}$ values based on a single average spectral energy distribution (SED; \citealt{ric06}) are usually appropriate when the disc is large compared to the innermost stable circular orbit (ISCO) around the black hole. While the overall size and surface area of a disc is set by its luminosity, for black holes of increasing mass the hole inside the ISCO grows proportionally to mass. Discs around higher-mass black holes thus lack the hottest and most UV-emitting parts, which are present in discs of similar luminosity around lower-mass black holes. This implies a non-linear dependence of $k_\mathrm{bol}$ on mass and disc luminosity, which was explored in the framework of thin-disc models by several authors \citep[e.g.][]{nem10,net19,lai23}. These explorations tend to offer a mean relation for $k_\mathrm{bol}$ as a function of monochromatic luminosity assuming a mean related black-hole mass or Eddington ratio.

For J0529, we use the 5100~\AA\ luminosity from Sec.~\ref{sec:lineprop} and the 33$^\circ$ inclination implied by our model. The mean relations by the three aforementioned works result in $k_\mathrm{bol}$ values ranging from 2.2 to 2.7, but are not applicable in the case of J0529. Such low $k_\mathrm{bol}$ values correspond to discs around black holes with masses of $>10^{10}~M_\odot$, whose SEDs visibly turn down towards the far-UV part of the spectrum. For example, \citet{lai23} show the red UV spectrum of their object J2157 that is in stark contrast to the blue spectrum of J0529, which implies a larger $k_\mathrm{bol}$ value for J0529 and suggests a lower-mass black hole for common stationary disc models. Also, using the mean relation from \citet{net19}, the luminosity of J0529 puts it far outside of the parameter space covered by the physically allowed range of black hole spin, while extrapolating their model calculation for any allowed spin to the correct luminosity puts it not only into a clearly super-Eddington regime but also suggests $k_\mathrm{bol}>10$.

Given the observed SED of J0529, $k_\mathrm{bol}\approx 2.5$ is clearly too low, while we also believe that the canonical average-SED case of $k_\mathrm{bol}=8.2$ from \citet{ric06} is presumably too high. This point cannot be settled without data on the intrinsic disc emission at $\lambda < 900$~nm. Using these corrections, we infer conservatively that $\log{L_\mathrm{bol}/\mathrm{erg\,s^{-1}}} = 47.74-48.28$ and hence $\lambda_\mathrm{Edd} = 6-19$ which is substantially super-Eddington.
In this regime, the luminosity increases only logarithmically with mass accretion rate \citep{abr88}, and there have been suggestions that the luminosity may saturate at a few to ten times $L_\mathrm{Edd}$ due to the decreasing radiative efficiency \citep{wat06,sad09}.
However, it might be possible to reach higher Eddington ratios: efficiencies that were unexpectedly high, exceeding those predicted by slim disc models, were reported by \cite{sad15} both for non-spinning, and even more so for spinning, black holes.
In the latter case, the authors suggested that rotational energy of the black hole was extracted by a process similar to the Blandford-Znajek mechanism, giving rise to the question of whether there would be an associated jet.
The only radio measurements available for J0529 is from the Rapid ASKAP Continuum Survey (RACS; \citealt{mcc20}), being 0.8~mJy at 0.9~GHz and 0.6~mJy at 1.4~GHz (D.~McConnell, priv. comm.).
Adopting the radio loudness parameter $R$ of \citet{kel89} as the ratio of the 6~cm flux density to that at 4400~\AA\ (both measured in mJy), we find an intermediate value of $R\sim14$ (which barely changes if one adjusts the rest-frame wavelengths to account for the typical redshift of $z\sim0.5$ of the \citealt{kel89} sample).
This transitional value suggests that J0529 has a jet, but also implies that the optical continuum remains dominated by thermal disk emission.

\cite{ina20} noted that the accuracy of the numerical algorithms plays an important role in the outcome of the simulations, with reference to calculations by \citet{jia14,jia19}.
In particular, \citet{jia19} state that at accretion rates of $\sim20-50~\dot{M}_\mathrm{Edd}$ the radiative efficiency remains high enough at $\gtrsim5$\% that one can have radiative luminosities of $10-20~L_\mathrm{Edd}$, but once the accretion rate reaches $\sim150~\dot{M}_\mathrm{Edd}$ the radiative efficiency drops to below 1\%.
The super-Eddington accretion also leads to a thick disk with large scale heights (e.g. $H/R \sim 0.7$, \citealt{jia19,jia25,hop25a}) consistent with what we find in our model.
Thus, J0529 may be characteristic of what is needed to grow massive black holes in the early universe \citep{ina20}: an extreme but physically plausible case of a QSO emitting close to its radiative limit in a regime where simulations indicate strong outflows are expected in a polar conical region.

\subsection{Outflow rate}
\label{sec:outflowrate}

Following \citet{shi19}, we estimate the outflow rate from our model as 
\[
\dot{M}_\mathrm{out} \ \propto \ \left( \frac{L_\mathrm{H\beta} \ V_\mathrm{out}} {n_e \ \delta_r} \right)
\]
where $\delta_r$ is the radial range over which the estimate is made.
We have applied this in a simplified way to calculate a single outflow rate between $R_\mathrm{min}$ and the mean radius $\mu_\mathrm{H\beta}$ at 0.9~pc.
For $V_\mathrm{out}$ we use the mean radial velocity of the clouds within this range, and we assume a canonical density of  $n_e = 10^{11}$~cm$^{-3}$.
The resulting outflow rate, based on the inferred \hb\ luminosity, is 
$\dot{M}_\mathrm{out} \sim 0.6$~M$_\odot$~yr$^{-1}$.
We can apply a factor 2 correction to compensate the anisotropy in the model and another factor 2 because of the mid-plane opacity to account for the reverse outflow, increasing the implied outflow rate to 
$\sim 2.4$~M$_\odot$~yr$^{-1}$.
This is very small compared to the accretion rate of 
$\dot{M}_\mathrm{acc} \gtrsim 200$~M$_\odot$~yr$^{-1}$ 
required to maintain the luminosity, assuming a radiative efficiency of $\eta = 0.05$ for super-Eddington sources \citep{jia19}.
However, it represents only the outflow of ionised gas in the BLR, and does not include the 
accretion disk wind launched on smaller scales, nor the radiation driven dusty wind from larger scales (which has the additional impact of reducing the inflowing mass that reaches the accretion disk).
The lower outflow rate associated with the BLR is consistent with what was found by \citet{vie18} for highly luminous QSOs. 
These authors showed that the kinetic power of the outflow from \civ\ in the BLR was an order of magnitude lower than that from \oiii\ on kiloparsec scales; and that winds from the accretion disk dominate the outflow on small scales to make up the difference.

\begin{figure*}[ht]
\sidecaption
\includegraphics[width=12cm]{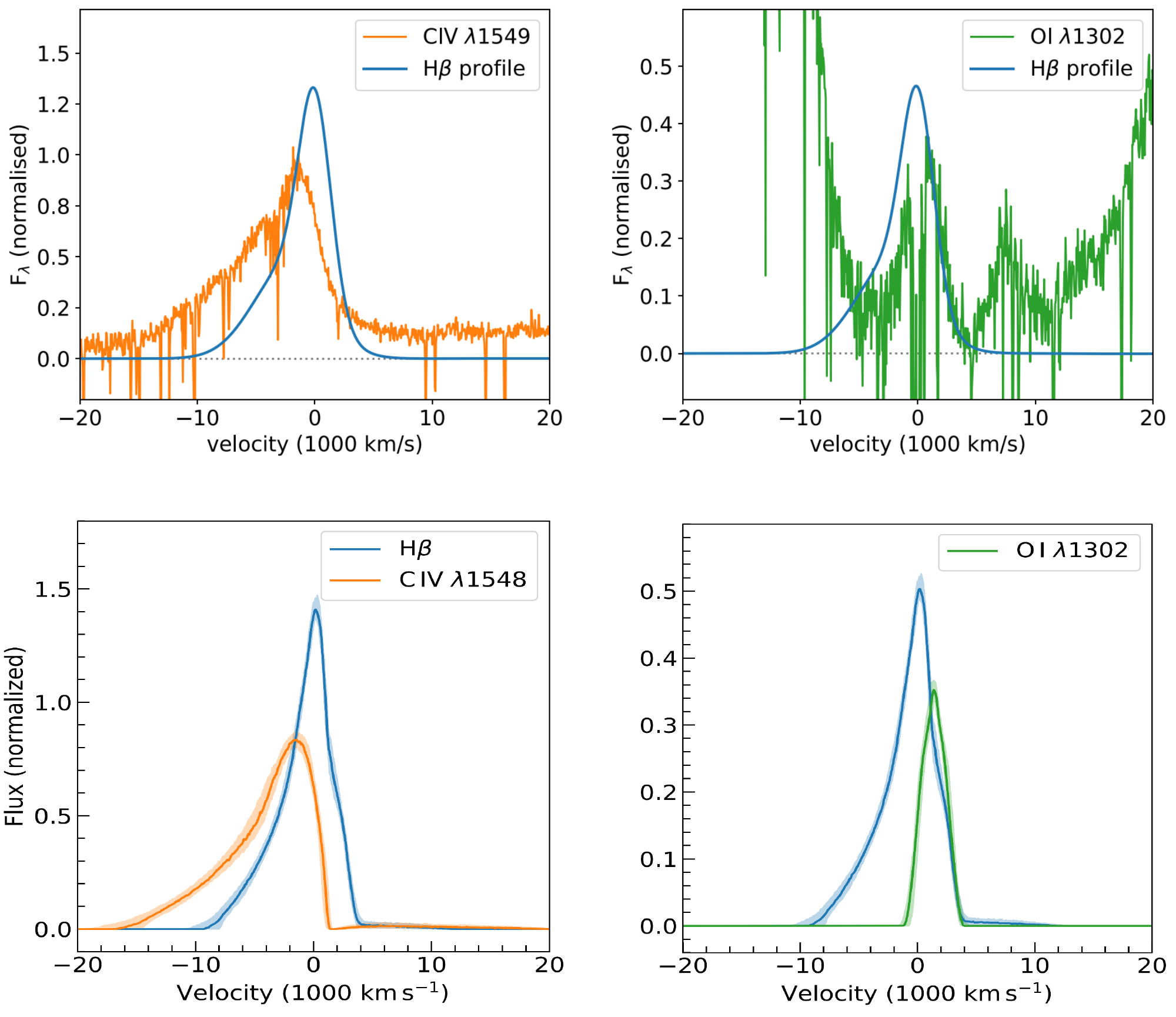}
\caption{Comparison of \hi\ profile to \civ\,$\lambda$1549~\AA\ and \oi\,$\lambda$1302~\AA.
The upper row shows the observed lines from the Xshooter spectrum \citep{wol24}. 
The velocity zero point here is set at the \hi\ line peak. The normalisation in each panel is adjusted to match at least part of the profile between the fitted \hi\ profile and the line profile in the plotted spectrum.
This highlights that the \civ\ is associated primarily with the BLR outflow and extends to higher velocities, while \oi\ is more associated with the static rotating part of the BLR (the source of the deep absorption at multiple velocities across the line peak is unclear.
The lower row shows the model predictions of the \civ\ (left) \oi\ (right) profiles (independent normalisations), with the velocity zero point also set at the \hi\ line peak. Details of how the predicted line profiles were created are given in Sec.~\ref{sec:civ_model} and~\ref{sec:oi_model}.}
\label{fig:linecomp}
\end{figure*}

\section{Comparison to the UV lines \civ\ and \oi}
\label{sec:linecomp}
\nolinenumbers

Differences between the high ionisation lines (HILs) and low ionisation lines (LILs) in the spectra of QSOs have been reported and analysed extensively in the literature for more than four decades.
In one of the first papers to address this topic, \citet{gas82} showed that the HILs were blue-shifted by typically 600~km~s$^{-1}$ with respect to the LILs, due to radial motions.
To explain this phenomenon, a physical model was proposed by \cite{col88} in which the LIL emission comes from clouds in the outer part of the accretion disk, while the HIL emission originates in outflowing transient clouds.
A better understanding of the physical situation came with higher quality spectra: wider band coverage, higher resolution, and better signal-to-noise.
In particular, detailed UV spectra of two local Narrow-Line Seyfert~1 AGN were presented by \cite{lei04a}.
These highlighted that the HILs were both strongly blue-shifted and asymmetric, while the LILs were narrow, symmetric, and at the systemic velocity of the AGN.
An extensive effort building on the eigenvector classification for local QSOs \citep{bor92} led to incorporation of the \civ\ line into the scheme, bringing it to 4-dimensions \citep{sul11,sul15, sul17}. 
It is now generally accepted that the line properties can be grouped into two classes of QSOs related to whether the \civ\ is dominated by a more systemic or strongly blue-shifted component. \citep{mar17,vie18,vie20,dec23}.

Despite this, \civ\ has been used to estimate $M_\mathrm{BH}$ because it is easily observable in high redshift QSOs and is known to originate in the BLR \citep{ves06,par13,run13}.
While this is physically reasonable for QSOs in which the line is both approximately symmetric and close to the systemic velocity, it is harder to explain a relation between the line properties and $M_\mathrm{BH}$ for emission originating in an outflow.
Nevertheless, \cite{coa17} found an empirical relation between the line's blue-shift and a correction factor that can be applied to the inferred mass, to match it to other scaling relations.
However, \cite{tem23} have shown that the situation is more complex, and that there is a 2-dimensional parameter space relating the blue-shift (and equivalent width) to a combination of $\lambda_\mathrm{Edd}$ and $M_\mathrm{BH}$.
They showed there is a wedge in that parameter space where the \civ\ blue-shift increases (and equivalent width decreases), and that in general the highest blue-shifts are observed in AGN with both $\lambda_\mathrm{Edd} \gtrsim 0.1$ and $\log{M_\mathrm{BH}/M_\odot} \gtrsim 9$.
Their modelling to reproduce these effects is bringing us to a better understanding of the cause of the two populations of QSOs with differing line properties noted above.

In Fig.~\ref{fig:linecomp} we compare the \hi\ line profile of J0529 to the \civ\ line, as well as the \oi\ in the Xshooter spectrum published by \citet{wol24}.
To do so, we have subtracted an estimate of the true continuum level as indicated by the lowest regions in the spectral ranges around the two lines.
We have not tried to fit the iron emission, nor other neighbouring emission, since the aim is only to compare the line profiles.

\subsection{Observation and Model of \civ}
\label{sec:civ_model}

The \civ\,$\lambda$1549~\AA\ line in the upper left panel of Fig.~\ref{fig:linecomp} appears to be mostly associated with the outflow, and can be traced to a velocity of $0.05c$.
It has a blue-shift of $\sim3600$~km~$^{-1}$, which is higher than the sample range of \cite{tem23} but consistent with their analysis based on the high $\lambda_\mathrm{Edd}$ and $M_\mathrm{BH}$, and confirms the nuclear nature of the outflow.

Using our BLR model we can make a simple prediction of the spectral profile of the line.
Because our model is based on dynamics rather than photoionisation, we can do this only in a crude way by adjusting the size of the BLR emission for the line according to an estimate of its expected radial location from RM measurements.
From the size-luminosity relation of the \civ\ line and 1350~\AA\ continuum (based on 7-13~yr RM campaigns of QSO samples) we estimate a characteristic radius for the \civ\ emission IN J0529 of 0.3~pc \citep{lir18,kas21,she24}.
We note, however, that there is large range of variation between relations including a possible dependence on luminosity;
and in addition the shape of the cross-correlation function may differ significantly between the \civ\ and \hi\ \citep{hor21}.

For this comparison, we have assumed that \civ\ arises only in the outflowing part of our model.
Although this is physically unlikely, the reason for doing so is that in the BLR model, both the disky and outflowing components extend over the same radial range with the same radial profile.
To modify the ratio of their contributions would require manually adjusting the scaling between them.
Although doing so would create a better match to the overall profile, it would not lead to further insights, and so we have chosen to focus on a comparison of the outflowing part.

The resulting line profile is shown in the lower left panel of Fig.~\ref{fig:linecomp}, where the stronger and more extended blue wing on the predicted \civ\ profile, which qualitatively matches the observation, is a result of the way that $V_\mathrm{out}(r)$ decreases with radius.
The higher excitation \civ\ originates from smaller radii and, lending support to the decreasing velocity profile we have adopted, has a higher outflow velocity.

\subsection{Observation and Model of \oi}
\label{sec:oi_model}

The upper right panel of Fig.~\ref{fig:linecomp} compares the \hi\ line profile to \oi\,$\lambda$1302~\AA, a low excitation line that in AGN is expected to arise in the BLR through Ly$\beta$ fluorescence \citep{net79,kwa81,mat07}.
\oi\,$\lambda$8446~\AA\ is also enhanced through this process and the ratio between these lines in the resulting radiative cascade can provide a measure of the extinction towards the BLR \citep{net79}.
Given that the Balmer line ratios in Sec.~\ref{sec:lineprop} hint that there might be some modest extinction, an independent measure like this would enable one to clarify the impact.
There is a good match on the red side of the profiles, but the \oi\ line shows less evidence for a blue wing suggesting it is more likely associated with the systemic velocity of the QSO.
We note that the peak of the \oi\ line is contaminated by deep absorption across a range of velocities, the cause of which is unclear.

With the same limitations and cautions as above for \civ, we can attempt to reproduce the \oi\ profile using our model of the BLR.
Because the emission is produced by Ly$\beta$ fluorescence, it is expected to arise at a similar location to the \hb\ line.
We have therefore set the radial scaling for the BLR to $\sim1$, noting that RM campaigns have shown that other low excitation lines such as \mgii\ and \feii\ also have broadly similar time lags to \hb\ \citep{mao93,hu15,pri23,she24}.
For simplicity we also assume that the \oi\ line originates only in the disky part of the BLR, although we acknowledge there is some indication of a weak outflow component in the observed profile.
The predicted \oi\ line is shown in the lower right panel of Fig.~\ref{fig:linecomp}.
In particular, the match of the red side of the profile tends to confirm that the emission is dominated by the disky part of the BLR; and, due to the difference of the blue side of the profile, that some emission must arise in the outflow.

\section{Conclusions}
\label{sec:conc}
\nolinenumbers

We have used the VLTI interferometer GRAVITY+ to spatially resolve the kinematics of the BLR of the luminous z=4 QSO SMSS~J052915.80-435152.0.
We measured the differential phase spectra across both the \hb\ and \hg\ lines and combined these with a detailed flux spectrum from ERIS.
We applied an updated modelling procedure that has a disky component and a separate outflowing component; and we fit both the flux profiles and phase spectra of the two \hi\ lines simultaneously, allowing only the radial profiles of their line emission distribution to vary.
While the redshift of the QSO inferred from the broad line peak is $z=3.962$, the systemic velocity for the best fit differs from that by $\sim1200$~km~s$^{-1}$ and favours a redshift of $z \sim 3.984$ within the scope of the model.
Based on the model, we conclude:
\begin{itemize}
\item 
The BLR is resolved to be about 1~pc in radius and is dominated by outflow, with velocities up to $\sim10^4$~km~s$^{-1}$ which create the prominent blue wing and also contribute to much of the line profile's core. Both high mid-plane opacity and anisotropic emission lead to the observed single-sided differential phase profile. Our model can plausibly predict both the observed blue-shifted \civ\,$\lambda$1549~\AA\ profile and the systemic \oi\,$\lambda$1302~\AA\ profile.
\item 
The black hole mass of $8\times10^{8}$~M$_\odot$ is the highest redshift measurement obtained directly from spatially resolved BLR kinematics to date. It is an order of magnitude lower than the masses inferred from a variety of scaling relations.
\item 
The QSO has a luminosity of $\sim6-19~L_\mathrm{Edd}$, which recent simulations indicate is plausible for accretion rates in a range around $20-50$~$\dot{M}_\mathrm{Edd}$ -- close to, but not yet, saturating the luminosity; and in a regime where strong polar conical outflows are expected.
\item
The ratio $\rfe =0.3$ (strength of the iron to \hb\ emission) would imply a rather modest accretion rate, contrary to other evidence indicating there is a high accretion rate.
We speculate that this discrepancy might be attributable to the dominance of the outflow in the BLR. If the scaling relations are applied using only the disky part of the BLR profile, the BH mass inferred from scaling relations is reduced, and the implied accretion rate increases. However, by comparing Fig.~\ref{fig:erisspec} and the bottom panels of Fig.~\ref{fig:avgphase} we caution that the decomposition of the BLR profile into outflowing and disky parts is dependent on geometry and anisotropy parameters, so cannot be done in a naive way by fitting a pair of Gaussian profiles.
\item
An estimate of the outflow rate of ionised gas in the BLR is $\sim 2.4$~M$_\odot$~yr$^{-1}$. This is about 1\% of the accretion rate, indicating it is only a minor part of the total outflow rate that includes winds launched on both smaller and larger scales.
\end{itemize}

\begin{acknowledgements}
GRAVITY+ is developed in a collaboration by the Max Planck Institute for extraterrestrial Physics, the Institute National des Sciences de l'Univers du CNRS (INSU) with its institutes LIRA~/~Paris Observatory-PSL, IPAG~/~Grenoble Observatory, Lagrange~/~C\^ote d’Azur Observatory and CRAL~/~Lyon Observatory, the Max Planck Institute for Astronomy, the University of Cologne, the CENTRA - Centro de Astrofisica e Gravita\c c\~ao, the University of Southampton, the Katholieke Universiteit Leuven, University College Dublin, Universidad Nacional Aut\'onoma de M\'exico and the European Southern Observatory. \newline
We thank the staff of Paranal Observatory for their enthusiasm and help during the visitor runs in which the observations here were performed.
This paper has benefitted from useful and lively discussion with many people on the BLR and outflows. 
In particular RD thanks B.~Vollmer and K.~Tristram, as well as the participants of the 2025 Vasto Accretion Meeting;
RD and TS thank H.~Netzer; 
JS thanks K. Inayoshi, J. Dai, and C. Jin for insightful suggestions on the BLR outflow modeling. 
JS is supported by The Fundamental Research Funds for the Central Universities, Peking University (7100604896) and the China Manned Space Program with grant no. CMS-CSST-2025-A09.
LCH was supported by the National Science Foundation of China (12233001) and the China Manned Space Program (CMS-CSST-2025-A09).
J.S-B. acknowledges the support received by the UNAM DGAPA-PAPIIT project AG-101025 and to the SECIHTI Ciencia de Frontera project CBF-2025-I-3033.
SH acknowledges support through UK Research and Innovation (UKRI) under the UK government’s Horizon Europe Funding Guarantee (EP/Z533920/1, selected in the 2023 ERC Advanced Grant round) and an STFC Small Award (ST/Y001656/1).
This work has been supported by the French Agence Nationale de al Recherche (ANR) through contracts AGN\_Melba ANR-21-CE31-001 and ExoVLTI ANR-21-CE31-0017.
We also thank the Fundação para a Ciência e Tecnologia (FCT), Portugal, for the financial support to the Center for Astrophysics and Gravitation (CENTRA/IST/ULisboa) through grant No. UID/PRR/00099/2025 (https://doi.org/10.54499/UID/PRR/00099/2025) and grant No. UID/00099/2025 (https://doi.org/10.54499/UID/00099/2025).
The authors thank D.~McConnell for providing the RACS measurements. These data were obtained from Inyarrimanha Ilgari Bundara / the Murchison Radio-astronomy Observatory. We acknowledge the Wajarri Yamaji People as the Traditional Owners and native title holders of the Observatory site. CSIRO’s ASKAP radio telescope is part of the Australia Telescope National Facility.
\end{acknowledgements}

\bibliographystyle{aa}
\bibliography{aa57285-25}

\begin{appendix}

\onecolumn
\section{Details of the model fit}
\label{app:details}
\nolinenumbers

The parameters for the best-fit model, and their 1-$\sigma$ ranges, are given in Table~\ref{tab:model}.
Here they have been grouped to make their role in the model easier to follow.
The model has been fitted to the \hb\ and \hg\ lines simultaneously.
As described in \cite{kuh24}, in practice this is done using two models in which all the geometry parameters as well as the global line emission parameters are tied to be the same in both.
The only parameters that can differ between the models are those listed specifically for the \hb\ versus \hg\ line properties.
These correspond to the shape parameter $\beta$, the radial scaling parameter $F$, and the peak flux scaling $F_\mathrm{peak}$.

\begin{figure*}[ht]
\centering
\includegraphics[width=0.7\hsize]{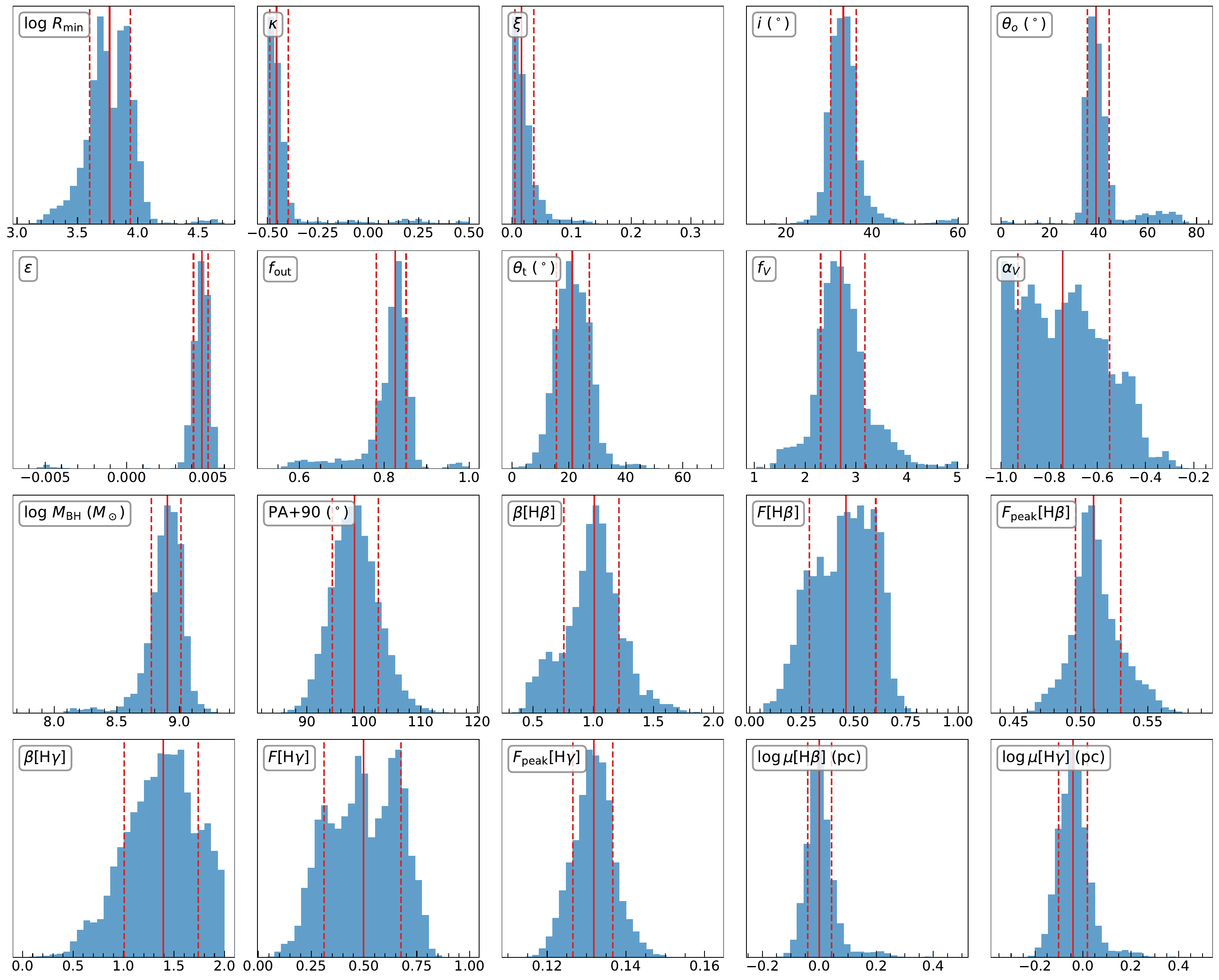}
\caption{Posterior histograms for the fitted parameters. The PA is shifted by $90^\circ$ for visual clarity.}
\label{fig:histograms}
\end{figure*}

\begin{figure*}[ht]
\centering
\includegraphics[width=0.6\hsize]{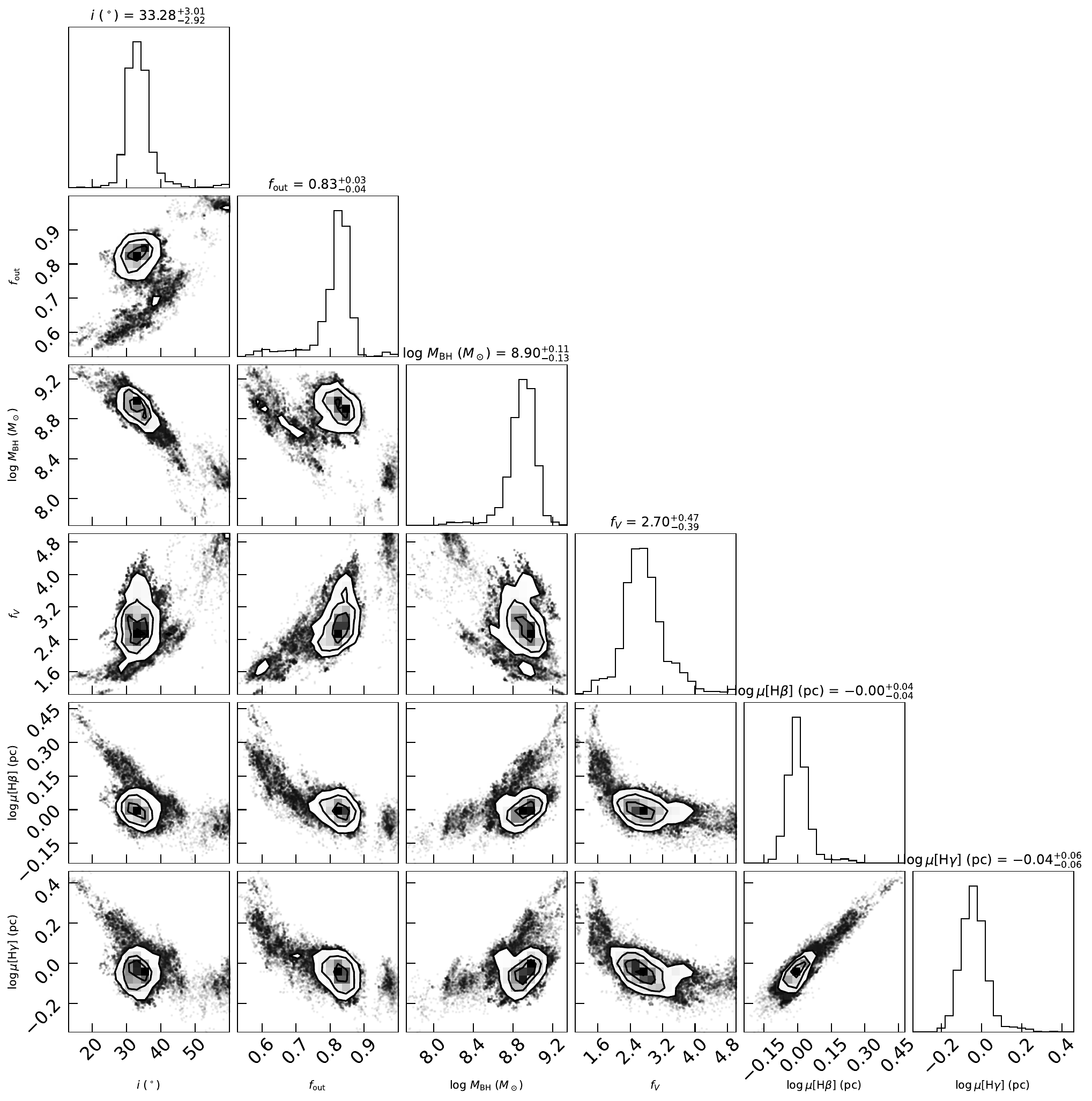}
\caption{Corner plot of a few key parameters. Four are fitted directly in the model and two are derived afterwards, propagating the uncertainties. The panels show there is a single major peak that is well constrained. They also show that at a low level of significance there are additional minor peaks and some expected correlations (e.g. that between M$_\mathrm{BH}$ and inclination is easiest to understand, and it leads to adjustment of other parameters).}
\label{fig:corner}
\end{figure*}

Histograms of the posterior distributions for the parameters are shown in Fig.~\ref{fig:histograms}, indicating that their ranges are generally well constrained by the data. 
As was done by \cite{kuh24}, we used the Python package 
\texttt{dynesty} \citep{spe2020} to fit the data with a nested sampling 
algorithm. Compared to the typical Markov Chain Monte Carlo algorithm, this 
method is powerful in terms of handling complex models with many (e.g. $>20$) 
parameters and a potentially multimodal posterior distribution.  We used the 
dynamic nested sampler (\texttt{DynamicNestedSampler}) which better estimates 
the likelihood function by re-sampling the posterior function a few more times 
after the ``baseline run.''  We used 1000 live points for the baseline run and 
added 500 points for each of the re-samplings until the default stop criteria was 
met.  We adopted the random walk algorithm (\texttt{rwalk}) to sample the prior 
space and used the multi-ellipse method (\texttt{multi}) to create the nest 
boundaries.  We adopted the default values for all the remaining options of 
\texttt{dynesty}. The reported fitting ends up with 2000 live points and 
$>20000$ effective samples, which is a statistically sufficient sampling of the 
posterior space.  A corner plot of selected parameters is presented in 
Fig.~\ref{fig:corner}.  In addition to the high-probability peak, some low-probability 
solutions appear due to the very detailed sampling.

\begin{figure*}[ht]
\centering
\includegraphics[width=1.0\hsize]{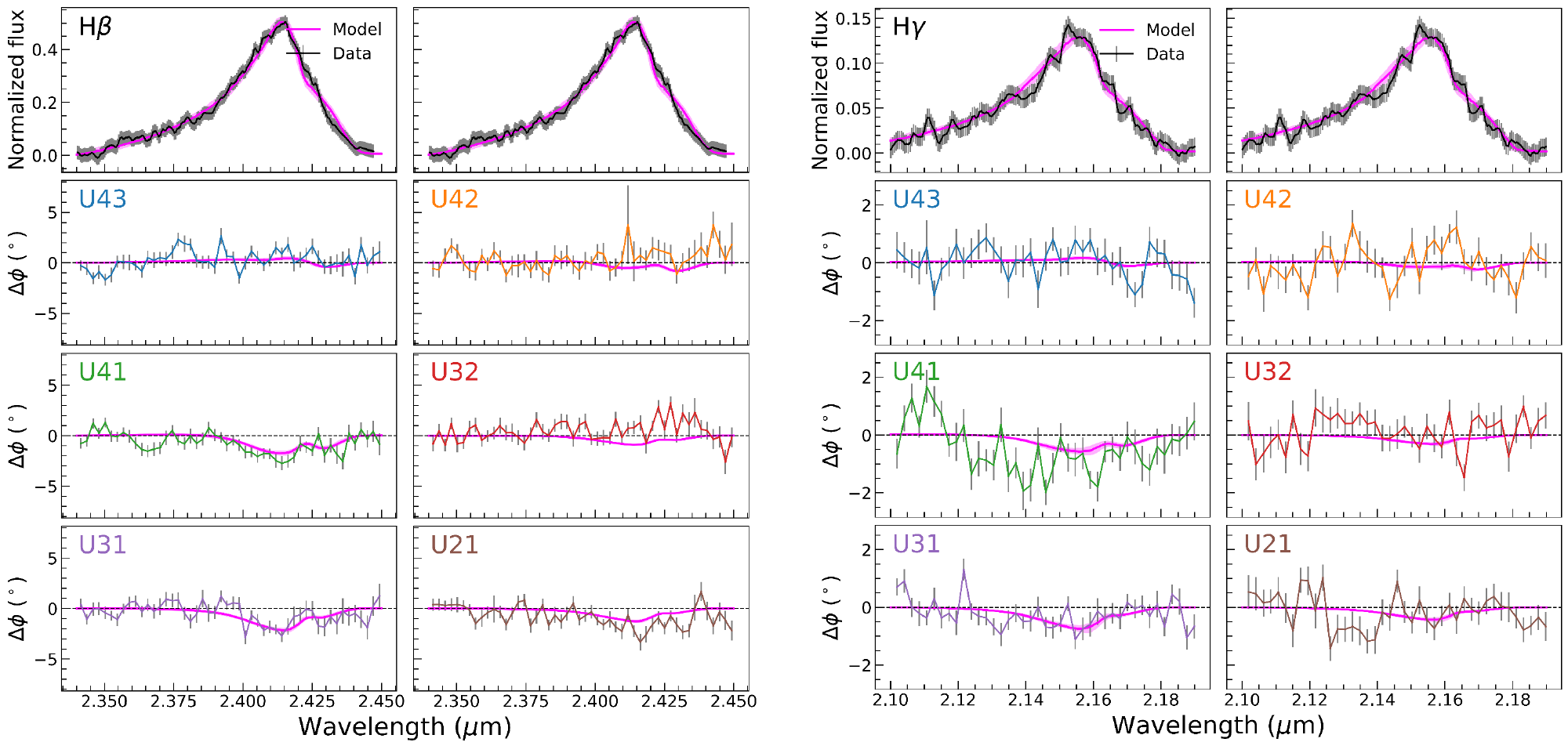}
\caption{Differential phase spectra for individual baselines. For the spectral segments around each of the \hb\ and \hg\ lines (top row), the measured and fitted differential phases are shown as a function of wavelength (other rows). The three baselines which contain the strongest signal (4-2, 4-1, and 3-1) are combined in Fig.~\ref{fig:avgphase}.}
\label{fig:allphases}
\end{figure*}

The resulting fits to the spectra and differential phases for the 6 baselines are shown in Fig.~\ref{fig:allphases}.
The strongest phase signals are seen for unit telescope (UT) baselines 4-1 and 3-1, which are the longest pair at 130~m and 102~m and differ in orientation by $<30^\circ$.
A weaker signal is present in baselines 3-2 and 2-1 which are at a very similar orientation to 3-1, differing by $<10^\circ$, and about half the length at 57~m and 47~m.
Very little signal is seen in the 4-3 and 4-2 baselines which are oriented in a different direction.
The phase signal has two peaks, the more significant one associated with the dominant outflow signature, and a weaker one related to the disky part of the BLR.

\begin{table*}[ht]
\caption{Model parameters and fitted values}
\label{tab:model}
\centering
\begin{tabular}{llllrc}
\hline \hline
& Parameter & Short description & Prior$^a$ & Value & 1~$\sigma$ range \\
\hline

\multicolumn{3}{l}{global geometry} \\
& $i$            & inclination (from face-on) [deg] & U($\cos{i}$: 0.5,1.0)$^b$ & 33.3 & -2.9,+3.0 \\
& $\mathrm{PA}$           & position angle of line of nodes (E of N) [deg] & U(-180,180) & 98.5 & -3.8,+4.4 \\
& $\log{M_\mathrm{BH}}$ & black hole mass (M$_\odot$) &U(7.0,11.0) & 8.90 & -0.13,+0.11 \\
& $f_\mathrm{out}$      & fraction of all the clouds in the outflowing part & U(0.0,1.0) & 0.83 & -0.04,+0.03 \\

\multicolumn{5}{l}{geometry of disky part} \\
& $\theta_0$  & angular extent above/below mid-plane [deg] &U(0,90) & 38.9 & -3.6,+5.3 \\

\multicolumn{5}{l}{geometry of outflowing part} \\
& $\theta_t$ & angular extent above/below disky part [deg] & U(0,90) & 21.1 & -5.6,+6.0 \\
& $f_V$      & scaling factor applied to $V_\mathrm{esc}$ at $R_\mathrm{min}$ to infer $V_\mathrm{out}(R_\mathrm{min})$ & U($\log{f_V}$: 0.0,0.7) & 2.70 & -0.39,+0.47 \\
& $\alpha_V$ & index denoting radial dependence of $V_\mathrm{out}(r)$ & U(-1.0,0.0) & -0.74 & -0.19,+0.19 \\ 

\multicolumn{5}{l}{global line emission properties} \\
& $\epsilon$      & shift of zero velocity wavelength, $\lambda_0 = (1+\epsilon) \lambda_\mathrm{initial}$ & U(-0.01,0.01) & 0.0046 & -0.0005,+0.0004 \\
& $R_\mathrm{min}$ & minimum radius of line emission [Schwarzschild radii] & U($\log{R_\mathrm{min}}$: 3.0,6.0) & 3.77 & -0.17,+0.17 \\
& $\kappa$        & line emission anisotropy (-0.5 to +0.5; 0 denotes isotropic) & U(-0.5,0.5) & -0.45 & -0.03,+0.06 \\
& $\xi$           & mid-plane opacity (0 to 1; 0 denotes opaque) & U(0.0,1.0) & 0.016 & -0.011,+0.021 \\

\multicolumn{5}{l}{\hb\ line properties} \\
& $\beta$[\hb]    & shape parameter for radial distribution of line emission & U(0.0,2.0) & 1.01 & -0.25,+0.20 \\
& $F$[\hb]        & scale factor such that mean radius $\mu = R_\mathrm{min} / F$ & U(0.0,1.0) & 0.46 & -0.18,+0.14 \\
& $F_\mathrm{peak}$[\hb] & flux scaling for peak of line profile & G(0.5,0.1) & 0.510 & -0.014,-+0.020 \\

\multicolumn{5}{l}{\hg\ line properties} \\
& $\beta$[\hg]    & shape parameter for radial distribution of line emission & U(0.0,2.0)& 1.39 & -0.39,+0.35 \\
& $F$[\hg]        & scale factor such that mean radius $\mu = R_\mathrm{min}/F$  & U(0.0,1.0) & 0.50 & -0.19,+0.18 \\
& $F_\mathrm{peak}$[\hg] & flux scaling for peak of line profile & G(0.14,0.02) & 0.132 & -0.005,+0.005 \\

\hline
\end{tabular}
\tablefoottext{a}{U refers to a uniform prior over the range given; G denotes a Gaussian prior with the centre and dispersion indicated.}
\tablefoottext{b}{Inclination is fitted as $\cos{i}$.}
\end{table*}

\end{appendix}

\end{document}